\documentclass[12pt,aps,amsmath,latexsym,amsfonts]{JHEP3}

\usepackage{epsfig}
\usepackage{amsfonts,amssymb,amsmath}
\usepackage[hang]{subfigure}
\usepackage{dsfont}

\newcommand{\be}{\begin{equation}}
\newcommand{\ee}{\end{equation}}
\newcommand{\bea}{\begin{eqnarray}}
\newcommand{\eea}{\end{eqnarray}}
\newcommand{\lsim}{\mbox{\raisebox{-.9ex}{~$\stackrel{\mbox{$<$}}{\sim}$~}}}

\def\ie{{\it i.e.}}
\def\nn{\nonumber}
\def\ni{\noindent}

\def\cF{{\mathcal F}}
\def\cA{{\mathcal A}}
\def\cM{{\mathcal M}}
\def\cN{{\mathcal N}}

\def\bcA{\boldsymbol{\mathcal A}}
\def\dbcA{\boldsymbol{\mathcal{\dot{A}}}}
\def\ddbcA{\boldsymbol{\mathcal{\ddot{A}}}}
\def\bk{\boldsymbol {\mathit k}}
\def\bx{\boldsymbol {\mathit x}}

\title{Statistical Anisotropy from Vector Curvaton in
D-brane Inflation}

\author{
Konstantinos
Dimopoulos$^a$\! \thanks{Email: konst.dimopoulos@lancaster.ac.uk} ,
Danielle Wills$^b$\! \thanks{Email: dwills@th.physik.uni-bonn.de} ,
Ivonne Zavala$^b$\! \thanks{Email: zavala@th.physik.uni-bonn.de}
\\
$^a$ Cosnortium for Fundamental Physics, 
Physics Department, Lancaster University, Lancaster LA1 4YB, UK. \\
$^b$ Bethe Center for Theoretical Physics and
Physikalisches Institut der Universit\"at Bonn,
Nu\ss allee 12, D-53115 Bonn, Germany. }

\abstract
{We investigate the possibility of embedding the vector curvaton paradigm in D-brane models of inflation in Type IIB string theory 
in a simple toy model.
The vector curvaton is identified with the U(1) gauge field that lives on the world volume of a D3-brane, which may be stationary or undergoing general motion in the internal space. The dilaton is considered as a spectator field which modulates the evolution of the vector field. In this set up, the vector curvaton is able to generate measurable
statistical anisotropy in the spectrum and bispectrum of the curvature perturbation assuming that the dilaton evolves as $e^{-\phi}\propto a^2$ where $a(t)$ is the scale factor. Our work constitutes a first step towards exploring how such distinctive features may arise from the presence of several light fields that naturally appear in string theory models of cosmology.
}


\begin{document}
\section{Introduction}
\subsection{Cosmic Inflation and the Vector Curvaton paradigm}

Ever-improving galaxy surveys have revealed to us that structure on the largest cosmological scales has an intricate net-like configuration, aptly named the Cosmic Web. According to the current model of structure formation, the gravitationally entangled strands that we see today have evolved from small density perturbations that left their imprint on the oldest source that is directly observed: the Cosmic Microwave Background (CMB). To understand how these ``seeds of structure'' came to be is an important challenge for cosmology.

The CMB also provides us with the strongest observational evidence for the homogeneity and isotropy of the universe on the largest scales, as it is uniform to about one part in a hundred thousand \cite{wmap}. We can then ask the question: What sort of processes or events in the earliest cosmological moments were able to make the universe so uniform, and yet still implant the tiny deviations from uniformity that could subsequently grow into the rich galaxy systems that we observe today?

The most successful framework for answering this question is the idea that the universe underwent a period of rapid, exponential or quasi-exponential expansion early on in its history, which on the one hand drove all classical perturbations to zero, but on the other hand was able to amplify fluctuations of the vacuum to set up the initial conditions for structure growth. This framework, dubbed cosmic inflation, also provides compelling insight into other seemingly unrelated problems in cosmology, such as the problem of the flatness of space, the problem of superhorizon correlations and the topological defects problem.

The amplification of vacuum fluctuations is of particular interest because, while the other problems can be overcome or at least ameliorated by \textit{any} models that are able to generate sufficiently long-lasting inflation, producing a spectrum of amplified fluctuations with the correct properties (as dictated by observations), necessitates the use of a more stringent approach. Such properties are manifest in the \textit{curvature perturbation} $\zeta$ which is generated by the dominant fluctuations, and typically includes a high degree of scale invariance, of Gaussianity and of statistical isotropy and homogeneity. As a first test, all viable models of inflation must be able to reproduce these features. However, precision measurements of the CMB have revealed the presence of finer-grained deviations from these basic properties which appear robust against foreground removal. For example, 
the latest observations favour a slightly red spectrum for $\zeta$, while there are hints of non-Gaussian features at 1-$\sigma$. Furthermore, 
the low multipoles of the CMB appear to be in alignment, which might imply the existence of a preferred direction on the microwave sky and therefore may constitute a violation of statistical isotropy \cite{AoE}. The capacity to provide a concrete explanation for the appearance of these deviations forms the finest sieve for models of inflation.

 Many inflation models assume that the energy density for inflating the universe and the vacuum fluctuations for seeding the galaxies must be provided by one and the same field, making it difficult to find a candidate field with all the right traits. There is however no \textit{a priori} reason to assume that this should be the case, given that the two jobs are rather independent. Indeed, any field that is around at the time of the expansion may end up having its vacuum fluctuations amplified, so long as it is sufficiently light and conformally non-invariant, regardless of which field is driving the expansion. If such a field is subsequently able to affect the expansion of the Universe then its own perturbation spectrum may be imprinted in the form of the curvature perturbation. One way for a field to do so is by dominating the energy density of the Universe after inflation, an idea that has been fruitfully expounded upon in the curvaton paradigm \cite{curv}. Such a field, which has nothing to do with expanding the Universe but instead does the job of generating the dominant contribution to $\zeta$, is then referred to as the \textit{curvaton} field.

Scalar fields, in addition to their wide use as inflatons, have also been popularly studied as curvatons, because if such fields are able to dominate the energy density of the Universe to imprint their spectrum, they will do so in an inherently isotropic way. A vector field on the other hand features an opposite sign for the pressure along its longitudinal direction relative to that along its transverse directions. Hence, if such a field (homogenised by inflation) were able to dominate the radiation background so as to imprint its own spectrum onto the curvature perturbation, it is likely to induce substantial anisotropic stress leading to excessive large-scale anisotropy when its pressure is non-vanishing, which is not observed. However, a massive homogeneous vector field undergoes coherent oscillations, much like a scalar field, and it has been shown that an oscillating vector field behaves like pressureless {\em isotropic} matter. Therefore, as long as the vector field begins to oscillate before its density parameter becomes significant, it may indeed dominate the radiation background in an observationally consistent way \cite{Dim1}. Thus, a vector field can, in principle, play the role of the curvaton.

The process by which the fluctuations of the vacuum are amplified in an expanding space is known as gravitational particle production, for the appearance of real particles in such a background is interpreted as the creation of particles by the changing gravitational field \cite{hawking}. A second problem which arises when using a vector field as a curvaton is related to this particle production process. A field must be sufficiently light to undergo gravitational particle production. However, a massless vector field is conformally invariant, therefore it does not respond to the expanding background and its vacuum fluctuations do not become amplified. 
This is an old problem which plagued the efforts to generate a primordial magnetic field during inflation. To overcome it many proposals introduced couplings to the vector field which break its conformality explicitly \cite{pmfinf}. A non-zero mass also breaks the conformal invariance of a vector boson field and was indeed employed in models of scalar electromagnetism to generate a primordial magnetic field \cite{scel}.

In Ref.~\cite{Dim1} the consequences of including a small non-zero vector mass were examined, and it was found that a vector field may indeed undergo particle production and obtain a scale invariant superhorizon spectrum of perturbations if the mass satisfies $m^2 \approx -2H_*^2$ during inflation, where $H_*$ gives the inflationary Hubble scale. This work pioneered the {\em vector curvaton} scenario, demonstrating for the first time that it is possible for the curvature perturbation to be affected or even generated by a vector field. The idea was implemented by coupling the vector field non-minimally to gravity through a $\frac16RA^2$ term in Refs.~\cite{nonmin,stanis}. However, this model was criticised for suffering from instabilities such as ghosts \cite{ghosts} (but see Ref.~\cite{RA2save}).

The vector curvaton scenario was further developed in Refs.~\cite{Dim1.1,Dim2}, where the supergravity motivated case of a vector field with a varying gauge kinetic function is explored as an alternative to demanding the afore-mentioned mass-squared condition, in order to avoid the issue of instabilities. There, it was shown that, under certain conditions, both the transverse and the longitudinal components of such a vector field can indeed obtain a scale invariant spectrum of superhorizon perturbations.

As demonstrated in Ref.~\cite{stanis} the contribution of a vector field to the curvature perturbation is in general statistically anisotropic. Thus, in the context of the vector curvaton mechanism, it then becomes possible to generate a statistically anisotropic contribution to the curvature perturbation, at once providing an explanation for the appearance of such a feature within a concrete paradigm (For a recent review see Ref.~\cite{VCrev}\footnote{In these works the vector field is assumed to be Abelian. For non-Abelian
vector fields and their potential contribution to $\zeta$ see Ref.~\cite{nonAbelian}.}). In the vector curvaton model of Refs.~\cite{Dim1.1,Dim2}, the gauge kinetic function and mass of the vector field are modulated by a scalar degree of freedom that varies during inflation, which may be the inflaton itself (e.g. see Ref.~\cite{lazajack} for a model realisation in supergravity). In the case that the statistical anisotropy in the spectrum generated by the vector field is large, the dominant contribution to $\zeta$ must come from an isotropic source, such as the scalar inflation. The vector curvaton then affects the curvature perturbation such that it acquires a measurable degree of statistical anisotropy.

\subsection{The D-brane Vector Curvaton}

 The idea that more than one field might have a role to play in the evolution of the early universe is highly motivated by string theory, which generically contains many fields, even at energies far below both the string and compactification scales. In standard inflation scenarios, the focus is often placed purely upon the behaviour of the candidate inflaton, while the other fields present in the set-up are considered to be either negligible or stabilised at the minima of their respective potentials. In the light of the curvaton scenario, it is interesting to consider how these fields might impact the evolution of the universe if they themselves undergo evolution. These fields appear with very precise couplings in the string set-up, and in this work we begin to explore how this precise structure dictated by string theory can enrich the cosmological picture. In particular, in open string D-brane models of inflation, the inflaton is typically identified with the position coordinate of a D$p$-brane moving in a warped throat \cite{InflaRev}. Such a brane features a world volume two form field $F_{MN}$, associated with the open strings that end on it. The components of such a field which correspond to Wilson lines have been  studied as potential inflaton candidates, both in the unwarped and warped  cases in Refs.~\cite{ACQ,AZ}.
 Thus it is  natural to investigate the role of the other components, in particular the 4D components, of such a field during the cosmological evolution, taking into account the precise way that these components couple to the various closed string modes. For example, these couplings can lead to a St\"uckelberg mass for the vector field. St\"uckelberg masses are ubiquitous in string theory and so it is interesting to consider whether these massive vector fields may give rise to cosmological signals, and one goal of the our work is to begin that line of enquiry. In addition, scalar moduli fields usually enter in the mass and gauge kinetic function for the vector field, and these can in principle vary during inflation. Such moduli include the dilaton as well as volume moduli for the case of wrapped branes. This indeed suggests the possibility  to embed the vector curvaton paradigm in D-brane models of inflation.

 In what follows we consider the simplest case of  vector fields on D3-branes, as a first attempt at demonstrating the vector curvaton mechanism in string theory, which may then be used as a starting point for constructing concrete cases. Vector fields on D3-branes may couple to the four-dimensional components of the bulk two-form $C_2$ which can lead to a St\"uckelberg mass for these fields. While it is true that in compactifications of Type IIB string theory with O3/O7-planes, as opposed to O5/O9-planes, the four-dimensional components of $C_2$ are projected out, we do not specify a compactification at this point, 
but aim at illustrating the mechanism rather than to provide a full model. We note however that for realistic compactifications with O3/O7-planes, one would need to consider D$(p>3)$-branes, in which case there is a non-vanishing 2-form suitable for generating a St\"uckelberg mass.

The prototypal scenario of slow-roll inflation has been described by the motion of a D3-brane along a particularly flat section of its potential in the throat, yielding a firm embedding for this scenario within a fundamental theory, as long as the potential can in fact support flat sections. Interestingly, further investigation of this picture within string theory led to an unexpected result: Inflation can take place even when the potential is very steep, leading to a completely new type of inflation known as Dirac-Born-Infeld (DBI) inflation \cite{st}. In DBI inflation, one no longer assumes that the kinetic term for the position field has its canonical form, but rather keeps the non-standard form just as it appears in the DBI action which describes the motion of the brane. This amounts to allowing the brane to move relativistically, bounded by the natural speed limit in the bulk, and the combination of this speed limit with strong warping in the throat allows for inflationary trajectories. In fact, for DBI inflation, achieving sufficient e-folds requires the potential to be \emph{very} steep rather than very flat.

In analogy to the DBI inflaton, while the U(1) vector field whose kinetic term appears in the DBI action is often considered to be of canonical form, keeping the general Born-Infeld form may lead to new features if this field plays a role in the cosmological evolution. A single vector field is unsuitable for the role of the inflaton, however the vector curvaton paradigm has demonstrated that it can indeed affect or even generate the curvature perturbation. In view of the fact that the vector curvaton paradigm considers, thus far, only the canonical vector field, part of this work is devoted to computing the equations of motion for the quantum fluctuations of the (inherently stringy) non-canonical field.

The majority of this work deals with the simpler case of a canonical vector field on a D3-brane, 
except Sections 3.1 and 3.2,
where the brane in question may either be stationary or undergoing general motion in the internal space. For this case, we provide a full cosmological analysis detailing how such a field may play the role of a vector curvaton and thus give rise to measurable statistical anisotropy in the spectrum and bispectrum of the curvature perturbation. As the dilaton appears in the St\"uckelberg mass and gauge kinetic function, we allow the possibility that it may undergo time evolution and thus treat it as a ``modulon'': a degree of freedom that varies during inflation and modulates the mass and gauge kinetic function of the vector field \cite{Dim2}. In order for the modulated vector field to give rise to a scale invariant spectrum, the gauge kinetic function and the mass must obey $f\propto a^{-1\pm 3}$ and $m\propto a$ respectively, where $a$ is the scale factor. In the case of the D3-brane vector field modulated by the dilaton, we found that these conditions are precisely met assuming $e^{-\phi}\propto a^2$ while the cosmological scales exit the horizon.
While this is promising from the cosmology point of view,
 it might be challenging  to realise concretely in string theory if the brane featuring the vector curvaton is static.
 This is because at tree-level the dilaton is expected to be stabilised at the energies in question. While it is plausible that the dilaton could be perturbed from its minimum, the effective potential would be quadratic, whereas the specified time evolution requires a linear potential. Away from the minimum, the potential for the dilaton is exponential, which is effectively linear for small displacements, therefore the required behaviour may be achievable if the dilaton is able to roll towards its minimum along these regions.
 On the other hand, for a moving brane in a generic warped background where the dilaton has a non-trivial profile, this will depend on the inflaton, and thus on time. In such a case, the required dependence on the scale factor might indeed be realised, but this will involve a careful study that goes beyond the scope of this work.
 In what follows we merely assume that this is possible.
 
 We point out however that we have chosen to use the dilaton as a modulon because it appears generically in the gauge kinetic function and mass for the D3-brane vector field, and, strikingly, with the precise powers in $f$ and $m$ that make it possible for the D3-brane vector field to generate a scale invariant spectrum: as we will discuss in what follows, if we impose $f \propto a^2$ then we automatically arrive at $m \propto a$, a non-trivial relation between the mass and gauge kinetic function that is specified by the vector curvaton paradigm, whereas it is not obvious at all that this relation should arise in string theory. Therefore, this simple picture nicely captures the key features of a stringy implementation of the vector curvaton paradigm, and we may readily compute the cosmological implications of these D-brane vector fields, to see how they may affect the curvature perturbation in the universe.

It should also be noted that in more complicated scenarios with branes of higher dimensionality, there may be other moduli which exhibit the appropriate powers in $f$ and $m$ that lead to scale invariance, and which are more attractive to use as modulons from the string theory point of view. Along these lines, the problems that arise from using the dilaton as a modulon are useful as guiding principles for selecting suitable moduli in realistic models.

Based on the assumption that the dilaton is able to evolve accordingly, we show that distinctive features in the curvature perturbation may arise from the intrinsic presence of several light fields in string theory models of cosmology, and notably from vector fields with St\"uckelberg masses. In particular, we discuss a scenario in which the dominant contribution to the curvature perturbation is given by the inflaton field, and the vector field can contribute measurable statistical anisotropy.
Throughout our paper we use natural units for which $c=\hbar=k_B=1$ and
Newton's gravitational constant is \mbox{$8\pi G=M_P^{-2}$}, with
\mbox{$M_P=2.4\times 10^{18}\,$GeV} being the reduced Planck mass. Sometimes we
revert to geometric units where \mbox{$M_P=1$}.

\section{The General Set-up}

In this section we discuss the set up which will be the basis for our investigation of the D-brane vector curvaton with non-standard kinetic terms.

We consider a  warped geometry in type IIB theory \cite{GKP,KS}. Thus, the ten dimensional metric takes the following form
(in the Einstein frame)
\be\label{metrica}
G_{MN}dx^Mdx^N=h^{-1/2}(y^A)\,g_{\mu\nu}\,dx^\mu dx^\nu
+ h^{1/2}(y^A)\,g_{AB} dy^A dy^B\,.
\ee
Here $h$ is the warp factor which depends on the compact coordinates $y^A$, and $ g_{AB}$ is the internal metric which may also depend on the compact coordinates.
This geometry is the result of having all types of fluxes present in the theory turned on: RR forms $F_{n+1}=dC_n$ for $n=0, 2, 4$ and their duals $n=6, 8$ (remember also that $F_5$ is self dual), as well as NSNS flux $H_3=dB_2$.
These fluxes have only  compact internal components, therefore their duals have legs in all four infinite dimensions plus the relevant components in the internal dimensions.

We now consider a probe D$p$-brane (or anti-brane) embedded  in this background. This has four of its dimensions lying parallel to the four  infinite dimensions, and $(p-3)$ spatial dimensions wrapped along an internal $(p-3)$-cycle. The dynamics of such a brane  is described by the sum of the Dirac-Born-Infeld (DBI) and  Wess-Zumino (WZ) actions.  The DBI part is given, in the Einstein frame, by\footnote{
In the string frame, the DBI action is given by
$$ S_{\rm DBI} = -\mu_p\,\int{d^{p+1}\xi\, e^{-\phi}
  \sqrt{- \det (\gamma_{ab}+ {\mathcal F}_{ab} )}}\,. $$
In D dimensions, the Einstein and string frames are related
by $G^E_{MN} = e^{-\frac{4}{(D-2)}\phi}G^s_{MN}$.  }
\be\label{dbi}
\!\,
S_{\rm DBI} = -\mu_p\,\int{d^{p+1}\xi\, e^{\frac{(p-3)}{4}\phi}
  \sqrt{- \det (\gamma_{ab}+ e^{-\frac{\phi}{2}}{\mathcal F}_{ab} )}}\,,
\hspace{-1cm}
\ee
where the tension of a D$p$-brane in the Einstein frame is\footnote{Notice that
for a D$3$-brane, $T_p=\mu_p$ in the Einstein frame.}
$$T_p = \mu_p \,e^{\frac{(p-3)}{4}\phi} \qquad
{\rm with} \qquad \mu_p = (2\pi)^{-p}(\alpha')^{-(p+1)/2}\,, $$
where $\phi$ is the dilaton, which in general depends also on $y^A$.
Furthermore, ${\mathcal F}_{ab}={\mathcal B}_{ ab}
+ 2\pi\alpha'\,F_{ab}$, with ${\mathcal B_2}$ the pullback of the NSNS background  2-form field  on the brane,
$F_2$ is the world volume gauge field we are interested in, and $\gamma_{ab} =  G_{MN}\,\partial_{a} x^M \partial_{b}  x^N$
is the pullback of the ten-dimensional metric on the brane in the Einstein frame.  Finally, $\alpha'=\ell_s^2$ is
the string scale and $\xi^a$ are the brane world-volume coordinates.
The indices $M,N = 0,1, ..., 9$; $a,b = 0, 1, ... ,p$; $\mu , \nu = 0,1, ..., 3$ and $A,B = 4, ..., 9$ label the coordinates in 10D spacetime, on the worldvolume, in 4D spacetime including the three extended dimensions, and in the remaining six internal dimensions respectively.
The action in Eq.~(\ref{dbi}) is reliable for arbitrary values of the gradients
$\partial_a  x^M$, so long as these are themselves
slowly varying in spacetime; that is, for small accelerations
compared to the string scale (equivalently, for small extrinsic
curvatures of the brane worldvolume). In addition, the string coupling
 should  be small in order for the perturbative expansion to hold, {\it i.e.}~$g_s\ll 1$, where $g_s = e^{\phi_0}$.

The Wess-Zumino  action describing the coupling of the D$p$-brane to the RR form fields is given by %
\be\label{wz}
S_{\rm WZ}= q\,\mu_p \,\int_{\mathcal W_{p+1}}\sum_n{\mathcal C}_n \wedge \,e^{\mathcal F}\,,
\ee
where ${\mathcal C}_n$ are the pullbacks of the background RR $C_n$ forms present, ${\mathcal F} = {\mathcal B} + 2\pi\alpha' F$ as before, and the  wedge product singles out the relevant terms in the exponential. Furthermore ${\mathcal W_{p+1}}$ is the world volume of the brane and $q= 1$ for a probe D$p$-brane, while  $q=-1$ for a probe D$p$-antibrane.

We now discuss these two actions in detail for our case of interest, a D3-brane.

\subsection{The Dirac-Born-Infeld Action}

Let us consider a probe D3-brane  positioned such that its three axes are aligned with the axes of the three extended spatial dimensions. We consider the brane to be moving in the internal space and so the internal coordinates become functions of the world volume coordinates, $y^{A}=y^{A}(\xi^\mu)$. In typical single-field inflation scenarios with D3-branes/$\overline{\rm{D3}}$-antibranes, the inflaton field is identified with the position coordinate of the D3-brane moving radially in the potential of the antibrane. Taking the pullback of the NSNS 2-form field $\mathcal{B}_2$ to vanish, the DBI action for a D3-brane is given by
\begin{align}
\label{DBI}
S_{\rm DBI} = -T_3 \int d^{4}x \sqrt{-{\rm det}(\gamma_{\mu\nu}+ e^{- \phi /2}\mathcal{F}_{\mu\nu})},
\end{align}
where
\be
\begin{array}{l}
\gamma_{\mu\nu} = h(r)^{-1/2}g_{\mu\nu}+ h(r)^{1/2}\partial_{\mu}y^{A}\partial_{\nu}y^{B}g_{AB},\\ 
\\
\mathcal{F}_{\mu\nu} = 2 \pi \alpha' F_{\mu\nu}\;,
\end{array}
\ee
with \mbox{$F_{\mu\nu}=\partial_\mu A_\nu-\partial_\nu A_\mu$}, and the warp factor depends only on the radial direction $r$.
We expand the determinant as follows:
\be
-{\rm det}[h^{-1/2}g_{\mu\nu}+ h^{1/2}\partial_{\mu}y^{A}\partial_{\nu}y_{A}+ 2 \pi \alpha' e^{- \phi /2}F_{\mu\nu}]=
-h^{-2}{\rm det}[g_{\mu\beta}]{\rm det}[\delta^{\beta}_{\ \nu}+ h\partial^{\beta}y^{A}\partial_{\nu}y_{A}+ lF^{\beta}_{\ \nu}]\,,
\ee
with
\begin{align*}
l = h^{1/2}2 \pi \alpha' e^{- \phi /2}.
\end{align*}
The DBI action in Eq.~(\ref{DBI}) then becomes
\be
\label{DBI2}
\!\,\begin{array}{l}
S_{\rm DBI} =
-T_3 \int d^{4}x h^{-1}\sqrt{-g}\sqrt{{\rm det}(\delta^{\beta}_{\ \nu}+ h\partial^{\beta}y^{A}\partial_{\nu}y_{A}+ lF^{\beta}_{\ \nu})}\,.
\end{array}\hspace{-1cm}
\ee

We now proceed to calculate the determinant:
\begin{align}
\label{determinant}
\begin{split}
&{\rm det}[\delta^{\beta}_{\ \nu}+ h\partial^{\beta}y^{A}\partial_{\nu}y_{A}+ lF^{\beta}_{\ \nu}]= 1 + \frac{l^{2}}{2}F_{\alpha\beta}F^{\alpha\beta}-\frac{l^4}{4}F_{\alpha\beta}F^{\beta\gamma}F_{\gamma\delta}F^{\delta\alpha} + \frac{l^4}{8}F_{\alpha\beta}F^{\alpha\beta}F_{\gamma\delta}F^{\gamma\delta}
\\
&+ hy_{A}^{\alpha}y^{A}_{\alpha}+ h^2y_{A}^{[\alpha}y^{A}_{\alpha}y_{B}^{\beta]}y^{B}_{\beta}+ h^3y_{A}^{[\alpha}y^{A}_{\alpha}y_{B}^{\beta}y^{B}_{\beta}y_{C}^{\gamma]}y^{C}_{\gamma}+ h^{4}y_{A}^{[\alpha}y^{A}_{\alpha}y_{B}^{\beta}y^{B}_{\beta}y_{C}^{\gamma}y^{C}_{\gamma}y_{D}^{\delta]}y^{D}_{\delta}
\\
&+3hl^2y_{A}^{[\alpha}y^{A}_{\alpha}F^{\beta}_{\ \gamma}F^{\gamma]}_{\ \beta}+ 3h^2ly_{A}^{[\alpha}y^{A}_{\alpha}y_{B}^{\gamma}y^{B}_{\beta}F^{\beta]}_{\ \gamma}+ 4h^3ly_{A}^{[\alpha}y^{A}_{\alpha}y_{B}^{\beta}y^{B}_{\beta}y_{C}^{\gamma}y^{C}_{\delta}F^{\delta]}_{\ \gamma}+ 4hl^3 y_{A}^{[\alpha}y^{A}_{\alpha}F^{\beta}_{\ \gamma}F^{\gamma}_{\ \delta}F^{\delta]}_{\ \beta}
\\
&+6h^2l^2y_{A}^{[\alpha}y^{A}_{\alpha}y_{B}^{\beta}y^{B}_{\beta}F^{\gamma}_{\ \delta}F^{\delta]}_{\ \gamma}\;,
\end{split}
\end{align}
where $y_{A}^{\alpha}\equiv \partial^{\alpha}y_{A}$ and the antisymmetrisation takes place over the Greek indices only.

\subsection{The Wess Zumino Action}

The general Wess Zumino (WZ) action for a D3-brane is given by (see Eq.~(\ref{wz}))
\be\label{wz3}
S_{\rm WZ}=
q \mu_{3}\int_{\mathcal{W}_{4}}  \left(\mathcal{C}_4+ \mathcal{C}_2 \wedge (2 \pi \alpha') F_2 +
\frac{ \mathcal{C}_0\, (2\pi \alpha')^2}{2} F_2 \wedge F_2\right)\,,
\ee
where we recall that the $\mathcal{C}_{n}$ are the pullbacks of the background $C_n$ forms  present in the flux background, $\mathcal{W}_{4}$
is the world volume, and $q$ gives the charge of the brane ($q=+1$ for a brane and $q = -1$ for an antibrane).

The first term in Eq.~(\ref{wz3}) simply gives the charge of the D3-brane.  In a flux compactification, the four form is given
by ${C}_4 = \sqrt{-g}\,  h^{-1}\, dx^0\wedge dx^1\wedge dx^2 \wedge dx^3$ and thus this term is essentially given by the warp factor.
The last term in Eq.~(\ref{wz3}) is the coupling of the axion field $C_0$ to the vector field we are interested in. In the present
case where the $C_0$ axion has been stabilised, this is just a topological term.\footnote{Were the axion not stabilised, the axial term might have been the source of parity violating statistical anisotropy as explored in Ref.~\cite{axial}.} Finally, the second term which couples the vector
field to the two-form, is a non-trivial term which is responsible for generating a mass for the $U(1)$ field via the St\"uckelberg
Mechanism, which is a standard mass generation mechanism in string theory according to which non-anomalous $U(1)$ vector fields may
acquire masses%
\footnote{As mentioned in the introduction, in a realistic type IIB flux compactification with O3/O7 planes, the four dimensional components of the RR-form $C_2$ are projected out. However, in what follows we stick with the St\"uckelberg mass mechanism, which is a ubiquitous mechanism in string theory and thus our work can readily be generalised to realistic models. In a concrete scenario, one could for example consider wrapped D5 or D7-branes in place of D3-branes. In such case there is a four dimensional 2-form that is not projected out by the orientifold action, and which couples to the vector field appropriately such that a mass is
generated via the St\"uckelberg mechanism. Thus for an explicit realisation, one would have to consider branes of different
dimensionality. On the other hand, one may stick with D3-branes but consider a more standard mass mechanism for such branes,
such as the Higgs mechanism. }. The details of how such a mass is generated are given in appendix~\ref{A1}.

 \subsection{The total 4-dimensional action}

 We may now write down the complete expression for the three fields we are considering. This includes the total action for the gauge field $A_\mu$ living on the D3-brane (containing the DBI and WZ pieces), as well as the actions for the position coordinate $r$ and the dilaton field $\phi$. Using the results in appendix \ref{A1} and considering the brane to be moving in the radial direction only (generalisation to multi-field scenarios will be commented upon later), the final D3-brane action, in terms of the canonically normalised vector field ${\mathcal A}_\mu=A_\mu/\tilde g$ with $\tilde g^2=T_3^{-1}(2\pi\alpha')^{-2}$ (see appendix A), is given by
\bea
\label{dbiwz3}%
\hspace{-1cm} S_{\rm D3} &=& -\int d^4 x \sqrt{-g}\Bigg(
  h^{-1}\sqrt{\Lambda}- qh^{-1}  + q \frac{m^2 }{2} {\mathcal A}_\mu  {\mathcal A}^\mu
- q  \frac{{\mathcal C}_0 }{8}\epsilon^{\mu\nu\lambda\beta} {\mathcal F}_{\mu\nu}{\mathcal F}_{\lambda\beta} \Bigg),   
\eea
where
\bea
&& \hspace{-1cm}
 \Lambda = 1+\frac{h e^{-\phi}}{2}  \cF_{\alpha \beta}\cF^{\alpha \beta} + \frac{h^2e^{-2\phi}}{8} \left({\mathcal F}_{\alpha \beta}
{\mathcal F}^{\alpha \beta} {\mathcal F}_{\gamma \delta}
{\mathcal F}^{\gamma \delta} - 2 {\mathcal F}_{\alpha \beta} {\mathcal F}^{\beta \gamma}{\mathcal F}_{\gamma \delta}
{\mathcal F}^{\delta \alpha} \right) + h\,\partial_\alpha\varphi\partial^\alpha \varphi
\nn \\ \nn \\
 && \hspace{5cm}
 +  3\,h\,l^2\left(\partial_\alpha \varphi\partial^\alpha \varphi {\mathcal F}_{\beta\gamma}{\mathcal F}^{\beta\gamma} - 2\partial_\alpha \varphi {\mathcal F}^{\alpha\beta}\partial^\gamma \varphi {\mathcal F}_{\gamma\beta}\right). \nn
\eea
In this expression we have introduced the canonically normalised  (fixed) position field defined by  $\varphi = \sqrt{T_3} \, r $ associated to the (radial) coordinate brane position $r$, and  the corresponding warp factor is defined  as $h(\varphi)  =  T_3^{-1} h(r)$.
The dilaton dependent mass is given in string units by
\begin{equation}
m^2 =e^{-\phi} (2\pi)^4\,\frac{M_s^2}{{\cal V}_6}\,, \label{m}
\end{equation}
where the dimensionless  (warped) 6D volume is defined as
$\mathcal{V}_6 = V_6 M_s^6 $ (see appendix~A).
 Furthermore  $\epsilon^{\mu\nu\alpha\beta}$ is the Levi-Civita tensor, such that $\epsilon_{0123} = \sqrt{-g}$.

 Coupling this action to four dimensional gravity, and including the necessary terms for an evolving dilaton as well as the potential for the brane's position,
 which will arise due to various effects such as fluxes  and presence of other objects, we can write\footnote{We drop the topological term at this stage since $C_0$ is stabilised and therefore we may use the fact that $\rm{dA \wedge dA = d(A \wedge dA)}$, hence this term constitutes a total derivative and thus will not give any effect in the cosmological evolution.}
\bea \label{total_action}
&&\hskip-1cm S=\frac{M_P^2}{2}\int d^4 x \sqrt{-g} \left[ R -\frac{1}{2}\partial_{\mu}\phi\,\partial^{\mu}\phi - V(\phi) \right]  \nn \\
&& \hskip3cm
-\int d^4 x \sqrt{-g}\left[  h^{-1}\sqrt{\Lambda} + V(\varphi)- qh^{-1}\!  + q \frac{m^2 }{2} {\mathcal A}_\mu  {\mathcal A}^\mu
\right],
\eea

\noindent where $M_P^2 = 2\,V_6 / [(2\pi)^7\alpha'^4]$ is the Planck mass.

\section{Stationary brane}

We are now ready to study the cosmological implications of the $U(1)$ gauge field which lives on a D3-brane world volume.
We start by considering the case in which the brane whose world volume hosts the vector field of interest is stationary in the internal space. Inflation is considered to be driven by a different D3-brane or any other working inflationary mechanism. Therefore the ``curvaton brane'' is just a D3-brane that may be present in the bulk at the time of inflation, for which \mbox{$V(\varphi)=\,$constant, $\dot\varphi$ =0.}
In what follows we look at two possibilities for the dilaton:  either it is fixed during inflation or it is able to evolve.

\smallskip

For a stationary D3-brane, the action in Eq.~(\ref{total_action}) then simplifies to

\bea\label{tot3}
 && \hskip-0.8cm S =  \frac{M_P^2}{2}\,\int d^4 x \, \sqrt{-g}\,  \left[ R -\frac{1}{2}\partial_{\mu}\phi\,\partial^{\mu}\phi - V(\phi) \right]\nonumber\\
&& \hskip-0.6cm
+ \!\! \int d^4 x \, \sqrt{-g} \Bigg\{ \! h^{-1}\!\!
\left[ q\! - \!\!\sqrt{1\!+\!\frac{h f(\phi)}{2}  {\mathcal F}_{\alpha \beta}{\mathcal F}^{\alpha \beta} + \frac{h^2f^2(\phi)}{8} \left({\mathcal F}_{\alpha \beta}
{\mathcal F}^{\alpha \beta} {\mathcal F}_{\gamma \delta}
{\mathcal F}^{\gamma \delta} - 2 {\mathcal F}_{\alpha \beta} {\mathcal F}^{\beta \gamma}{\mathcal F}_{\gamma \delta}
{\mathcal F}^{\delta \alpha} \right)} \right]
\nonumber \\
&&\hskip 11.5cm
-  q\, \frac{m^2(\phi)}{2} \, {\mathcal A}_\mu  {\mathcal A}^\mu
\Bigg\} \,,
\eea
where we have defined 
\be
f(\phi) = e^{-\phi} \qquad {\rm and } \qquad 
m^2(\phi) = e^{-\phi } \tilde m^2 \,.
\ee
We now focus on the  equations of motion for the vector field, derived from the action above. These are given by (from now on  we take $q=1$)
\bea
G_{\mu\nu}&=&
  \frac{\gamma_A}{2}\left(
  f\,{\mathcal F}_{\mu\beta}{\mathcal F}_{\nu}^{\ \beta}
 + \frac{h f^2\, }{2}\,{\mathcal F}_{\mu\kappa}
 {\mathcal F}_{\nu}^{\ \kappa}   {\mathcal F}^2
  - h  f^2 \,{\mathcal F}_{\nu}^{\ \sigma}{\mathcal F}_{\sigma\delta}
 {\mathcal F}^{\delta\kappa}{\mathcal F}_{\kappa\mu}\right)\nn  \\
&& \hskip2cm
 + \frac{m^2 e^{\phi}}{2}{\mathcal  A}_{\mu}{\mathcal A}_{\nu}
+ \frac{g_{\mu\nu}\, h^{-1}}{2}
\left(   1-   \gamma_A^{-1} -\frac{m^2 }{2}
  \,{\mathcal A}_\mu  {\mathcal A}^\mu     \right), \\ \nn \\
 \sqrt{-g}\, m^2 {\mathcal A}^{\nu}&=&
 \partial_{\mu}\left[ \sqrt{-g} \, \gamma_A
 \left( f\,{\mathcal F}^{\mu\nu}
 - h f^2 \,{\mathcal M}^{\nu\mu}
 +\frac{h f^2}{2}\,{\mathcal N}^{\mu\nu}  \right)\right], \label{Aeom}
\eea
where we have defined:
\be
\gamma_A^{-1}\equiv
\sqrt{1+\frac{hf}{2} \, {\mathcal F}^2
 + \frac{h^2f^2}{8} \left( {\mathcal F}^4 - 2 {\mathcal F}_{\alpha \beta} {\mathcal F}^{\beta \gamma}{\mathcal F}_{\gamma \delta}
{\mathcal F}^{\delta \alpha} \right)}\,,
\ee
\bea
 &&{\mathcal M}^{\nu\mu} = {\mathcal F}^{\nu\beta}{\mathcal F}_{\beta\gamma} {\mathcal F}^{\gamma\mu} \,,\label{calM}\\
&& {\mathcal N}^{\mu\nu}={\mathcal F}^{\mu\nu}
 {\mathcal F}_{\alpha\beta}{\mathcal F}^{\alpha\beta}
\label{calN}
\eea

\ni and $G_{\mu\nu} = R_{\mu\nu}-\frac{g_{\mu\nu}}{2}R$ is the Einstein tensor.
In a FRW universe, the four dimensional metric takes the usual form
\be
ds^2 = - dt^2 + a^2(t) \delta_{ij} \, dx^idx^j,
\ee
where $a(t)$ is the scale factor. Moreover, we can expect inflation to homogenise the vector field, and therefore for the background solution
\be
\partial_i \cA_{\mu} =0\,. 
\ee

Using this condition,  one can check that the factor $\gamma_A$ associated with the vector field is given simply by
\be
\label{Agamma}
\gamma_A = \frac{1}{\sqrt{1+\frac{h f }{2} \cF^2}} =
\frac{1}{\sqrt{1- h f a^{-2} \dbcA \cdot \dbcA }} \;.
\ee
Moreover, the $\nu=t$
component of the vector field equation implies that
\be
\cA_t =0 \,,
\ee
and we thus have $\cA_\mu = (0,{\boldsymbol{\mathcal A}}(t))$.
Using this, the $\nu = i $ component of the equation of motion becomes
\be
\label{backgroundA}
\gamma_A \ddbcA +\gamma_A \dbcA\left(H+\frac{\dot\gamma_A}{\gamma_A} + \frac{\dot f}{f} \right) + \frac{m^2}{f}\bcA =0\,,
\ee
where $H \equiv \frac{\dot a}{a} $ is the Hubble parameter
and we have used the fact that the $\cM$ and $\cN$ terms cancel each other in the background solution.\\
\\
From the form of Eq.~(\ref{backgroundA}) we see that the effective mass of the vector $M\equiv\frac{m}{\sqrt{f}}$ is constant and given by
\be
\label{physmass}
M= \sqrt{\pi}\,(2 \pi)^{5}\, \frac{M_P}{\mathcal{V}_6}\,.
\ee

\smallskip

\begin{figure}
\centering
\includegraphics[scale=0.50]{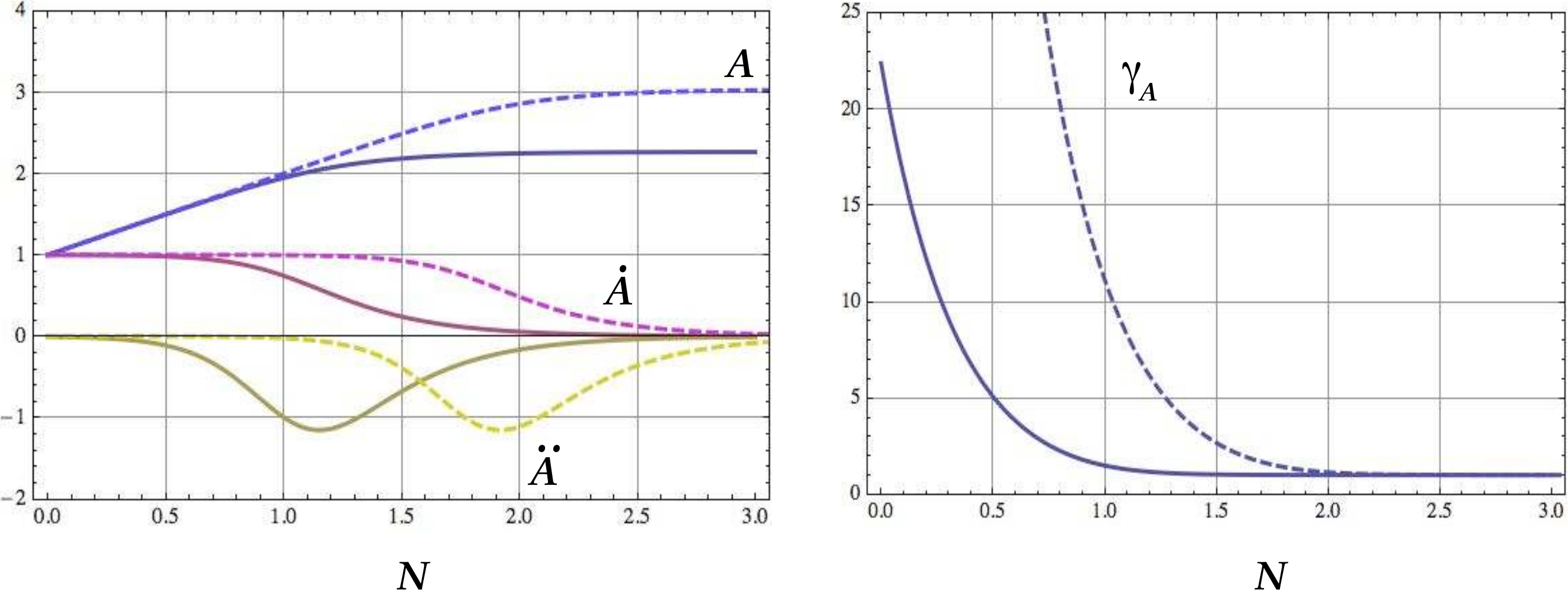}
\caption{The qualitative behaviour for the vector background
with an evolving dilaton.
The solid lines  correspond to initial conditions on the vector field such that $\gamma_{A_{ini}} \sim 22$ and the dashed lines to $\gamma_{A_{ini}}\sim 224$. 
The horizontal axis measures the elapsing e-folds.}
\label{fig. 1}
\end{figure}

As we demonstrate below, the desired behaviour of our system is attained when
\mbox{$f\propto a^2$}.
Solving Eqs.~(\ref{backgroundA}) and (\ref{Agamma}) in the case that $M \ll H$ and $e^{-\phi} \propto a^2(t)$, we obtain the behaviour of the background $\cA_\mu$ and $\gamma_A$ during the inflationary 
period. The results are plotted in Fig.~\ref{fig. 1}, in which we indicate the qualitative behaviour of the vector background and its time derivatives as well as that of $\gamma_A$. The solid lines correspond to initial conditions on the vector field such that $\gamma_{A_{ini}} \sim 22$ and the dashed lines to $\gamma_{A_{ini}}\sim 224$. The horizontal line indicates the number of e-folds elapsed, $N$.
We see that the background soon freezes out at constant amplitude during inflation, while $\gamma_A$ converges very quickly to 1. This is in agreement with the findings of 
Ref.~\cite{Dim2}, when \mbox{$f\propto a^2$}.

\subsection{Perturbations in the general case}

We now  calculate the perturbations of the vector field during the cosmological evolution to see what sort of new terms arise in the most general case of varying dilaton and non-standard vector kinetic terms. 

We perturb the vector field around the homogeneous value $\cA_\mu(t)$ as follows:
\be\!\,
\begin{array}{l}
\cA_\mu(t,{\bf x}) =\cA_\mu(t)+\delta \cA_\mu(t,{\bf x}) \quad \Rightarrow\\
\\
\cA_t(t,{\bf x}) =  \delta\cA_t(t,{\bf x})  \quad \& \quad
\bcA(t,{\bf x}) = \bcA(t) +  \delta\bcA(t,{\bf x})\,.
\end{array}\hspace{-1cm}
\ee

Then, the equation of motion for the temporal component $\nu =t$ becomes
\be\label{tcomp}
a^2m^2\delta\cA_t + f \gamma_A \left[ \nabla\cdot \delta\dbcA-\nabla^2\delta\cA_t \right] + hf^2a^{-2}\gamma_A^3 \dbcA\cdot
\nabla\left[\dbcA \cdot \delta\dbcA- \dbcA\cdot \nabla(\delta\cA_t)\right] =0\,.
\ee

\smallskip

The spatial component takes a more complicated form. Combining it with the integrability condition: $\partial_{\nu} $(\ref{Aeom}), gives:
\bea\label{icomp}
&& \hskip-0.5cm
-am^2\delta\bcA -a f \gamma_A \! \left(\!\delta \ddbcA + \left[ H +
 \frac{\dot\gamma_A}{\gamma_A} + \frac{\dot f}{f}  \right] \delta\dbcA   
 -a^{-2}\nabla^2\delta\bcA 
 - \!\left[ \frac{\dot\gamma_A}{\gamma_A} + \frac{\dot f}{f}  - 2H - 2\frac{\dot m}{m}  \right] \!\nabla\delta \cA_t \!\right)
  \nn \\ \nn\\
&& \hskip0.3cm   - h f^2 a^{-1}\gamma_A^3\dbcA \,
\dbcA \cdot \left(\delta \ddbcA  +\left[ -H+ 3\frac{\dot\gamma_A}{\gamma_A} +2\frac{\dot f}{f} \right]   \delta\dbcA \right)
 \nn \\ \nn\\&&\hskip0.3cm
  - hf^2 a^{-1}\gamma_A^{3}
\left\{\dbcA (\ddbcA\cdot\delta\dbcA)+ \ddbcA (\dbcA\cdot\delta\dbcA)
-\dbcA[\ddbcA \cdot\nabla(\delta \cA_t)] -\ddbcA[\dbcA\cdot\nabla(\delta\cA_t)]  \right\} \nn\\ \nn\\
&&\hskip0.3cm   + hf^2 a^{-1}\gamma_A^3\dbcA(\dbcA\cdot\nabla)
\left( \left[ 3\frac{\dot\gamma_A}{\gamma_A} -4H + 2\frac{\dot f}{f}-2\frac{\dot m}{m}\right] \delta\cA_t 
			+ a^{-2}\nabla\cdot\delta\bcA\right) \nn\\ \nn\\
&&\hskip0.3cm   +h f^2\gamma_A a^{-3}
\Big\{-\dbcA(\dbcA\cdot\nabla)(\nabla\cdot\delta\bcA)  +
\dbcA \,\nabla^2(\dbcA\cdot\delta\bcA)    - \dbcA\cdot\nabla [\nabla(\dbcA\cdot \delta \bcA)]   \nn\\   \nn\\
&&\hskip0.3cm + (\dbcA\cdot\nabla)(\dbcA\cdot\nabla)\delta\bcA  - (\dbcA\cdot\dbcA)\nabla^2\delta\bcA  + (\dbcA\cdot\dbcA)
\nabla(\nabla\cdot \delta \bcA)\Big\} =0 \,.
\eea

We now pass to momentum space by performing a Fourier expansion of the perturbations as follows:

\be\label{fourier}
\delta\cA_\mu (t,\boldsymbol{x}) = \int{\frac{d^3k}{(2\pi)^{3/2}} \delta\cA_\mu(t,\bk) \exp{(i\bk\cdot\bx})}\,.
\ee

Plugging this into (\ref{tcomp}) and (\ref{icomp}), we can write the equations of motion for the transverse and longitudinal components as follows. Making the definitions: 
$$ \delta \boldsymbol{\mathcal{A}}^{||} \equiv \frac{\boldsymbol{\mathit{k}}(\boldsymbol{\mathit{k}}\cdot \delta \boldsymbol{\mathcal{A}})}{{k}^2},
 \hspace{1cm}
\hspace{1cm}
\delta \boldsymbol{\mathcal{A}}^{\bot} \equiv \delta \boldsymbol{\mathcal{A}} - \delta \boldsymbol{\mathcal{A}}^{||}\,, $$
\bea%
k^2 \equiv \bk \cdot \bk \,,\qquad  \qquad
\gamma_A = \frac{1}{\sqrt{1-h f \left(\frac{\boldsymbol{\mathcal{\dot{A}}}}{a}\right)^2}}  \,, \nn
\eea%
\bea
 {\bf Q} &=& \frac{hf \gamma_A^2}{a^2}  \left[\boldsymbol{\mathcal{\dot{A}}}(\boldsymbol{\mathcal{\ddot{A}}}\cdot \boldsymbol{\mathit{k}}) + \boldsymbol{\mathcal{\ddot{A}}}(\boldsymbol{\mathcal{\dot{A}}}\cdot \boldsymbol{\mathit{k}})+ \boldsymbol{\mathcal{\dot{A}}}(\boldsymbol{\mathcal{\dot{A}}}\cdot \boldsymbol{\mathit{k}})\left(3\frac{\dot{\gamma_A}}{\gamma_A} + 2\frac{\dot f}{f}- 4H - 2\frac{\dot m}{m}\right)\right]  \nn \\
&&\hskip8cm  + \boldsymbol{\mathit{k}}\left(\frac{\dot{\gamma_A}}{\gamma_A} + \frac{\dot f}{f} - 2H - \frac{\dot m}{m}\right)\,, \nn \\ \nn
\\
R & =&  k^2 + \frac{(am)^2}{\gamma_A f}  + \frac{h f \gamma_A^2}{a^2} (\boldsymbol{\mathcal{\dot{A}}}\cdot \boldsymbol{\mathit{k}})^2\,,
\nn
\eea
the transverse component becomes

\begin{align}\label{trans1}
&\left\{ \partial_{t}^2 + \left(H + \frac{\dot{\gamma_A}}{\gamma_A}
+ \frac{\dot f}{f}\right)\partial_{t} + \frac{m^2}{\gamma_A f} + \left(\frac{k}{a}\right)^2 + \frac{h f}{a^2}\left[\left(\frac{\boldsymbol{\mathcal{\dot{A}}}\cdot \boldsymbol{\mathit{k}}}{a}\right)^2- \boldsymbol{\mathcal{\dot{A}}}^2\left(\frac{k}{a}\right)^2\right] \right\}\delta \boldsymbol{\mathcal{A}}^{\bot} \nn \\\nn
\\ \nn
& + h f\left(\frac{\gamma_A}{a}\right)^2\left\{ \boldsymbol{\mathcal{\ddot{A}}} +  \left(\frac{3\dot{\gamma_A}}{\gamma_A}+\frac{2\dot f}{f} 
- H\right)  \boldsymbol{\mathcal{\dot{A}}}   
+ \frac{\bf Q}{R} (\boldsymbol{\mathcal{\dot{A}}}\cdot \boldsymbol{\mathit{k}})\right\} \boldsymbol{\mathcal{\dot{A}}}\cdot \delta \boldsymbol{\mathcal{\dot{A}}}^{\bot} \nn\\\nn \\
&  +\frac{hf}{a^4} \left[\boldsymbol{\mathcal{\dot{A}}}\,k^2 - (\boldsymbol{\mathcal{\dot{A}}}\cdot \boldsymbol{\mathit{k}})\boldsymbol{\mathit{k}}\right]\boldsymbol{\mathcal{\dot{A}}} \cdot \delta \boldsymbol{\mathcal{A}}^{\bot} + hf \left(\frac{\gamma_A}{a}\right)^2 
\left\{\boldsymbol{\mathcal{\dot{A}}}(\boldsymbol{\mathcal{\dot{A}}}\cdot \delta \boldsymbol{\mathcal{\ddot{A}}}^{\bot}) +
\boldsymbol{\mathcal{\dot{A}}}(\boldsymbol{\mathcal{\ddot{A}}}\cdot \delta \boldsymbol{\mathcal{\dot{A}}}^{\bot}) \right\}= 0 \,,
\end{align}

\bigskip

\ni while the longitudinal component is:

\begin{align}\label{long1}
&\hskip-0.7cm
\left\{ \!\partial_{t}^2 \!+ \!\left[H + \frac{\dot{\gamma_A}}{\gamma_A}+ \frac{\dot f}{f} + \frac{1}{R}\left(\! 2H + \frac{2\dot m }{m} -\frac{\dot{\gamma_A}}{\gamma_A} -\frac{\dot f}{f} \right)k^2\right]\! \partial_{t}+\frac{m^2}{\gamma_A f } +\! \left(\frac{k}{a}\right)^2 \!\!+ \frac{h f}{a^2}\left(\frac{\boldsymbol{\mathcal{\dot{A}}}\cdot \boldsymbol{\mathit{k}}}{a}\right)^2\!\right\}\delta \boldsymbol{\mathcal{A}}^{||} \nn\\
\nn  \\ \nn
& \hskip-0.4cm
+ \frac{h f}{a^2} \left[\gamma_A^2\left( \frac{k}{a}\right)^2 \!\! \boldsymbol{\mathcal{\dot{A}}} - \frac{(\boldsymbol{\mathcal{\dot{A}}}\cdot \boldsymbol{\mathit{k}})\boldsymbol{\mathit{k}}}{a^2}\right]\boldsymbol{\mathcal{\dot{A}}} \cdot \delta \boldsymbol{\mathcal{A}}^{||} 
+h f \left(\frac{\gamma_A}{a}\right)^2\left\{ \left[ 1-\frac{k^2}{R}  \right]\boldsymbol{\mathcal{\ddot{A}}} - \frac{\bf Q}{R} (\boldsymbol{\mathcal{\dot{A}}}\cdot \boldsymbol{\mathit{k}}) \right. \nn\\ \nn \\  
& \left. 
+ \left[\frac{3\dot{\gamma_A}}{\gamma_A} +\frac{2\dot f}{f} - H - \left(\frac{ 3\dot{\gamma_A}}{\gamma_A} + \frac{ 2\dot f}{f}- 4H -\frac{2\dot m}{m}\right) \frac{k^2}{R} \right] \boldsymbol{\mathcal{\dot{A}}} 
\right\} \boldsymbol{\mathcal{\dot{A}}}\cdot \delta \boldsymbol{\mathcal{\dot{A}}}^{||} 
 \nn \\ \nn \\ 
   &  +  hf\left(\frac{\gamma_A}{a}\right)^2\left\{
   \boldsymbol{\mathcal{\dot{A}}}(\boldsymbol{\mathcal{\dot{A}}}\cdot \delta \boldsymbol{\mathcal{\ddot{A}}}^{||}) + 
\left[1- \frac{k^2}{R}\right] \boldsymbol{\mathcal{\dot{A}}}  (\boldsymbol{\mathcal{\ddot{A}}}\cdot \delta \boldsymbol{\mathcal{\dot{A}}}^{||} ) \right\}
= 0  \,.
\end{align}

\bigskip

At this point we can compare these equations to the standard kinetic term case \cite{Dim1.1,Dim2}. In fact, the first lines in equations (\ref{trans1}) and (\ref{long1}) have the same form as the standard  kinetic term case for the vector field with $f \cF^2$. However in the present case, we obtain extra terms coming from the vector `Lorentz' factor $\gamma_A$. 
This factor acts in a way analogous to the gauge kinetic function, but it is independent of the dilaton field.
Moreover the Lorentz factor also adds second order terms in the background, coming from $\cM, \cN$ in the equations of motion [cf. Eqs.~(\ref{calM}) and (\ref{calN})]. These terms cannot be neglected once we consider $\gamma_A >1$.

\subsection{Non standard kinetic term, $\gamma_A>1$}

In order to have some insight into Eqs.~(\ref{long1}) and (\ref{trans1}) we consider the various projections of the perturbations onto the vector background. First, we decompose the background into components parallel and perpendicular to the momentum vectors:

\be
\bcA = \bcA^{||} + \bcA^{\bot}.
\ee
For the perturbations $\delta\bcA = \delta\bcA^{||} + \delta\bcA^{\bot}$ we define
\bea\label{alpha beta}
&\delta\bcA^{\bot} \equiv \delta\bcA^{\bot_{||}}  + \delta\bcA^{\bot_{\bot}} 
\eea
such that
\bea
&\delta\bcA^{\bot_{||}} \cdot \bcA^{\bot} = \delta\cA^{\bot_{||}}  \cA^{\bot}\nn\\ \nn
& \delta\bcA^{\bot_{\bot}}  \cdot \bcA^{\bot} = 0.
\eea
In this way, $\delta\bcA^{\bot_{||}} $ and $\delta\bcA^{\bot_{\bot}} $ are the modes of $\delta\bcA^{\bot}$ that are parallel and perpendicular to $\bcA^{\bot}$ respectively.

Now taking $\bk \cdot$(Eq.~(\ref{trans1})) and using the decomposition in Eq.~(\ref{alpha beta}) we obtain an equation for $\delta\bcA^{\bot_{||}} $,

\bea\label{alpha1}
\delta \ddot\cA^{\bot_{||}} = - \left[\frac{\ddot\cA^{\bot}}{\dot\cA^{\bot}}+ \frac{\ddot\cA^{||}}{\dot\cA^{||}}+ 3\frac{\dot\gamma_A}{\gamma_A}+ 2\frac{\dot{f}}{f} - H + \frac{Q^{||}}{R}k\right] \delta\dot\cA^{\bot_{||}} \,.
\eea
where $Q^{||} = \bk \cdot {\bf Q}$.
So we see that this mode is not oscillating on any scales and therefore it will not give rise to particle production.

To obtain an equation for $\delta\bcA^{\bot_{\bot}} $, we now take $\delta\bcA^{\bot_{\bot}} \cdot$(Eq.~(\ref{trans1})) which yields
\bea\label{beta}
\!\!\!\delta\ddot\cA^{\bot_{\bot}}  + 
\!\left(\!H + \frac{\dot\gamma_A}{\gamma_A} + \frac{\dot{f}}{f}\right)\!\delta\dot\cA^{\bot_{\bot}}  +\!\left\{\!\frac{m^2}{\gamma_A f} +\!\left(\frac{k}{a}\right)^2\!\left[1- \frac{hf}{a^2}\left((\dot\cA^{\bot})^2+ 2\dot\cA^{\bot}\dot\cA^{||}\right)\!\right]\!\!\right\}\delta\cA^{\bot_{\bot}}  =0 \,.\nn\\
\eea
Taking the small-scale limit of Eq.~(\ref{beta}), we see that we have particle production at early times and the Bunch-Davies vacuum is well-defined, as soon as the mass term becomes negligible in the UV limit. However the sound speed is reduced compared to the canonical oscillator. This equation does not give rise to instabilities because $\gamma_A^{-2}$ is always positive. 

Now we turn to the equation for the longitudinal component $\delta\bcA^{||}$. Taking $\bk \cdot$(Eq.~(\ref{long1})) and using the decomposition in Eq.~(\ref{alpha beta}) we obtain

\bea\label{rho}
&\delta\ddot\cA^{||} + \left\{H + \frac{\dot\gamma_A}{\gamma_A}+ \frac{\dot{f}}{f} + \frac{k^2}{R}\left(2H + 2\frac{\dot{m}}{m}- \frac{\dot\gamma_A}{\gamma_A} - \frac{\dot{f}}{f}\right)+ 2hf\left(\frac{\gamma_A}{a}\right)^2\left(1- \frac{k^2}{R}\right)\ddot\cA^{||}\dot\cA^{||}\right.\nn \\ \nn
\\ & \left.+ hf\left(\frac{\gamma_A}{a}\right)^2(\dot\cA^{||})^2\left[- \frac{Q^{||}}{R}+ 3\frac{\dot\gamma_A}{\gamma_A}+ \frac{\dot{f}}{f}- H - \frac{k^2}{R}\left(3\frac{\dot\gamma_A}{A}+ \frac{\dot f}{f}\right)\right]\right\}\frac{\delta\dot\cA^{||}}{B} \nn \\ \nn
\\& + \left[\frac{m^2}{B\gamma_A f}+\left(\frac{k}{a}\right)^2\right]\delta\cA^{||} = 0,
\eea
where
\be
B = 1 + hf\left(\frac{\gamma_A}{a}\right)^2 (\dot\cA^{||})^2.
\ee
Taking the small-scale limit we obtain a canonical oscillator equation,
\be
\delta\ddot\cA^{||} +\left(\frac{k}{a}\right)^2\delta\cA^{||} = 0\,,
\ee
therefore the Bunch-Davies vacuum is well-defined and we once again have particle production at early times. Therefore we see that particle production can take place for two of the three modes only, namely in the $\delta\bcA^{\bot_\bot}$ and $\delta\bcA^{||}$ directions, and hence is inherently anisotropic. After inflation, both of these modes will oscillate due to the mass term in Eqs.~(\ref{beta}) and (\ref{rho}) and could affect the curvature perturbation via the vector curvaton mechanism. Note in particular that the mass terms are suppressed by $\gamma_A$ and therefore these modes are indeed effectively massless at early times while $\gamma_A$ is large.

Taking the late-time limit of these equations, \emph{i.e.} $\gamma_A \rightarrow 1$ and $\dot\cA \rightarrow$ const., we obtain   \bea\label{beta late}
\delta\ddot\cA^{\bot_\bot} + \left(H +  \frac{\dot{f}}{f}\right)\delta\dot\cA^{\bot_\bot}  +\left[\frac{m^2}{f} +\left(\frac{k}{a}\right)^2\right]\delta\cA^{\bot_\bot}  =0,
\eea
and
\be\label{rho late}
\delta\ddot\cA^{||} + \left[H +  \frac{\dot{f}}{f} + \frac{\left(\frac{k}{a}\right)^2}{\left(\frac{k}{a}\right)^2 +\frac{m^2}{f}}\left(2H + 2\frac{\dot{m}}{m}-  \frac{\dot{f}}{f}\right)\right]\delta\dot\cA^{||} = 0.
\ee
As we will see in the following section, these equations give rise to exactly scale-invariant superhorizon spectra as long as $f \propto a^2$ while the cosmological scales exit the horizon. However, as a consequence of the intermediate regimes in Eqs.~(\ref{beta}) and (\ref{rho}), it is not clear that the resultant spectra will be scale-invariant. Due to the fact that the regime in which $\gamma_A \neq 1$ is short relative to the regime in which $\gamma_A = 1$, this could add some scale-dependence to the spectrum.

 We can get further inside into the behaviour of the perturbations by looking at the their superhorizon limits. In order to see this cleare, we pass to conformal time and make some useful definitions:
\bea
&& \int \frac{dt}{a(t)} = \tau, \qquad ' = d/d\tau , \qquad {\cal H} = \frac{a'}{a},
\qquad 
\alpha = \frac{1}{\sqrt{f \gamma_A}}, \qquad z= \alpha^{-1},\\\nn\\
&& \delta {\boldsymbol {\cal A}} =  \delta {\boldsymbol {\cal W}}/\sqrt{f \gamma_A}, \\\nn\\
&& R = k^2\left[ 1+ \frac{(a m)^2}{k^2 \gamma_A f} + \frac{h f \gamma^2_A}{ k^2 a^4}
({\boldsymbol {\cal A'}}\cdot {\boldsymbol{\mathit{k}}})\right] = k^2(1+r), \\\nn\\
&& \bar {\bf Q} = \frac{hf \gamma_A^2}{a^4}  \left[\boldsymbol{\mathcal{{A'}}}(\boldsymbol{\mathcal{{A''}}}\cdot \boldsymbol{\mathit{k}}) + \boldsymbol{\mathcal{{A''}}}(\boldsymbol{\mathcal{{A'}}}\cdot \boldsymbol{\mathit{k}})+ \boldsymbol{\mathcal{{A'}}}(\boldsymbol{\mathcal{{A'}}}\cdot \boldsymbol{\mathit{k}})\left(3\frac{{\gamma_A'}}{\gamma_A} + 2\frac{ f'}{f}- 6{\cal H} - 2\frac{ m'}{m}\right)\right] \nn  \\
&&\hskip8cm  + \boldsymbol{\mathit{k}}\left(\frac{{\gamma_A'}}{\gamma_A} + \frac{ f'}{f} -
2{\cal H} - \frac{ m'}{m}\right)\,.
\eea

\ni Using this information, the new equations for the perturbations take the form:

\begin{align}\label{transW}
&\delta \boldsymbol{\mathcal{W''}}^{\bot} + \left[ \frac{m^2 a^2}{\gamma_A f} + k^2\left( 1 -
\frac{h f}{a^4} (\boldsymbol{\mathcal{{A'}}})^2 \sin^2{\theta_{\boldsymbol{\cal A}\boldsymbol {\mathit{k}}}}\right) -\frac{z''}{z} \right] \delta \boldsymbol{\mathcal{W}}^{\bot}
+\frac{hf}{a^4} k^2 \boldsymbol{\mathcal{A'}}^{\bot} \left( \boldsymbol{\mathcal{A'}}^{\bot}\cdot \delta\boldsymbol{\mathcal{W}}^{\bot} \right)
 \nn \\\nn
\\ \nn
&\hskip1cm + \frac{h f\gamma^2_A}{a^4}\left\{ \left[\boldsymbol{\mathcal{{A''}}} +  \left(3\frac{{\gamma_A'}}{\gamma_A}+2\frac{ f'}{f}
-4 {\cal H}+2\frac{\alpha'}{\alpha}\right)  \boldsymbol{\mathcal{{A'}}}
+ \frac{\bf {\bar Q}}{R} (\boldsymbol{\mathcal{{A'}}}\cdot \boldsymbol{\mathit{k}})\right] \boldsymbol{\mathcal{{A'}}}\cdot \delta \boldsymbol{\mathcal{{W'}}}^{\bot}
\right. \nn\\ \nn \\
& \hskip3cm\left.
+  \boldsymbol{\mathcal{{A'}}} \left[ \frac{\alpha'}{\alpha}\boldsymbol{\mathcal{{A''}}}\cdot \delta \boldsymbol{\mathcal{{W}}}^{\bot} + \boldsymbol{\mathcal{{A''}}}\cdot
\delta \boldsymbol{\mathcal{{W'}}}^{\bot} + \boldsymbol{\mathcal{{A'}}}\cdot
 \delta \boldsymbol{\mathcal{{W''}}}^{\bot} \right]
\right. \nn\\ \nn \\
& \hskip-0.1cm \left.
+ \left[ \frac{\alpha'}{\alpha} \boldsymbol{\mathcal{{A''}}} + \left(\frac{\alpha''}{\alpha} + \frac{\alpha'}{\alpha}\left( \frac{3{\gamma_A'}}{\gamma_A}+\frac{2 f'}{f}
-4 {\cal H}\right) \right)  \boldsymbol{\mathcal{{A'}}}   + \frac{\bar {\bf Q}}{R}\frac{\alpha'}{\alpha}  (\boldsymbol{\mathcal{{A'}}}\cdot \boldsymbol{\mathit{k}}) \right]
\boldsymbol{\mathcal{{A'}}}\cdot  \delta \boldsymbol{\mathcal{{W}}}^{\bot}
\right\} =0\,,
\end{align}

\ni while the longitudinal component is:

\begin{align}\label{longW}
&\delta \boldsymbol{\mathcal{W''}}^{||} + \frac{2}{1+r}\left({\cal H} -\frac{m'}{m} + \frac{\alpha'}{\alpha}\right) \delta \boldsymbol{\mathcal{W'}}^{||}
+\frac{hf\, k^2 }{a^4} \left( \gamma^2_A \boldsymbol{\mathcal{A'}} - \boldsymbol{\mathcal{A'}}^{||}\right) \boldsymbol{\mathcal{A'}}  \cdot \delta\boldsymbol{\mathcal{W}}^{||}  \nn \\  \nn \\ \nn
& + \left[\frac{2}{1+r}\left({\cal H} -\frac{m'}{m}+ \frac{\alpha'}{\alpha} \right)\frac{\alpha'}{\alpha}  - \frac{z''}{z} + \frac{m^2 a^2}{\gamma_A f} + k^2\left( 1 +
\frac{h f}{a^4} (\boldsymbol{\mathcal{{A'}}})^2 \cos^2{\theta_{\boldsymbol{\cal A}\boldsymbol {\mathit{k}}}}\right) \right] \delta \boldsymbol{\mathcal{W}}^{||}
 \nn \\\nn
\\ \nn
&\hskip-1cm
 + \frac{h f\gamma^2_A}{a^4}\left\{\!\! \Bigg[ \!\left(\frac{r}{1+r}\!\right)\!\boldsymbol{\mathcal{{A''}}} +  \left(\frac{3{\gamma_A'}}{\gamma_A}+\frac{2 f'}{f}
-4 {\cal H} +\frac{2\alpha'}{\alpha}  -\frac{1}{1+r} \left( \frac{3{\gamma_A'}}{\gamma_A}+\frac{ 2f'}{f}
-6 {\cal H} -\frac{2 m'}{m}\right)\right)  \boldsymbol{\mathcal{{A'}}}
\right. \nn\\ \nn \\
& \hskip2cm\left.
- \frac{\bf {\bar Q}}{R} (\boldsymbol{\mathcal{{A'}}}\cdot \boldsymbol{\mathit{k}})\Bigg] \boldsymbol{\mathcal{{A'}}}\cdot  \delta \boldsymbol{\mathcal{{W'}}}^{||}
+ \boldsymbol{\mathcal{{A'}}} \Bigg[ \boldsymbol{\mathcal{{A'}}}\cdot
 \delta \boldsymbol{\mathcal{{W''}}}^{||} + \boldsymbol{\mathcal{{A''}}}\cdot
 \delta \boldsymbol{\mathcal{{W'}}}^{||} \left(1-\frac{1}{1+r}\right) \Bigg]
\right. \nn\\ \nn \\
& \hskip-0.7cm\left.
+ \Bigg[ \!\!\left(1-\frac{1}{1+r}\right) \frac{\alpha'}{\alpha} \boldsymbol{\mathcal{{A''}}} + \! \left(\!\frac{\alpha''}{\alpha} + \frac{\alpha'}{\alpha}
\left[\frac{3{\gamma_A'}}{\gamma_A}+\frac{ 2f'}{f}
-4 {\cal H}   -\frac{1}{1+r}\! \left[ \frac{3{\gamma_A'}}{\gamma_A}
+\frac{ 2f'}{f} -6 {\cal H} -\frac{2 m'}{m}\right]\right] \!\right) \boldsymbol{\mathcal{{A'}}}
\right. \nn\\ \nn \\
& \hskip1cm \left.
- \frac{\bf {\bar Q}}{R} (\boldsymbol{\mathcal{{A'}}}\cdot \boldsymbol{\mathit{k}}) \frac{\alpha'}{\alpha}\Bigg]  \boldsymbol{\mathcal{{A'}}}\cdot \delta \boldsymbol{\mathcal{{W}}}^{||}   +
\frac{\alpha'}{\alpha} \left(1-\frac{1}{1+r}\right) \boldsymbol{\mathcal{{A'}}}
\left(\boldsymbol{\mathcal{{A''}}}\cdot
\delta \boldsymbol{\mathcal{{W}}}^{||} \right)\Bigg]
\right\} =0 \,.
\end{align}

Taking the limit $k/aH \to 0$ we see that the perturbations evolve in exactly the same way. 
Therefore from the analysis above we see that on one hand that particle production will be anisotropic as expected and moreover, the perturbations will evolve in the same way in the superhorizon limit. 
Thus, the statistical anisotropy will be preserved but, as noted above, scale invariance is not-guaranteed.
In the next subsection, we study in detail the perturbations when the kinetic term for the vector field takes a standard form. 

\subsection{Standard kinetic term, $\gamma_A\sim 1$}
\label{subsec:varying_dilaton}

In this section we focus on in the case in which $\gamma_A \sim 1$.
We begin by outlining the conditions which should be placed on the evolution of the dilaton in order to achieve a scale-invariant spectrum of vector perturbations. It was clearly demonstrated in Ref.~\cite{Dim2} that, for a light vector field evolving during quasi-de Sitter expansion with a gauge kinetic function $f \propto a^\alpha$ and mass $m \propto a^\beta$, the power spectrum of vector perturbations is exactly scale invariant when $\alpha = -1\pm3$ and $\beta = 1$, as long as the vector field remains light when the cosmological scales exit the horizon. In the case that $f\propto a^{-4}$ (and $m \propto a$), the power spectra for the transverse and longitudinal components of the vector perturbations can become roughly equal if the vector field becomes heavy by the end of inflation. This allows for (approximately) isotropic particle production and entails that the vector field can provide the dominant contribution to the curvature perturbation, which is known to be predominantly isotropic. In the other case, {\it i.e.}~$f \propto a^2$ (and $m \propto a$), the power spectrum for the longitudinal  component of the vector perturbations is much larger than that of the transverse component, therefore particle production is strongly anisotropic and the dominant contribution to the curvature perturbation must come from some other, isotropic source, {\it e.g.}~a scalar field.\footnote{This is so even for the case in which \mbox{$f\propto a^{-4}$} (and \mbox{$m\propto a$}) if the vector field remains light until the end of inflation \cite{Dim2}.} However, the important point is that, in this case, the vector field can contribute measurable statistical anisotropy.

In our scenario, $f \propto e^{-\phi}$ and $m \propto e^{-\phi /2}= \sqrt{f}$ [cf. Eqs.~(\ref{m})], such that if we demand $m \propto a$ we must have $f \propto a^2$. %
This suggests an explicit realisation of the latter of the above two possibilities, which leads to scale invariance of the vector perturbations as shown in Ref.~\cite{Dim2}, as long the D-brane vector field is light and evolving during quasi-de Sitter expansion, and 
 $e^{-\phi}\propto a^{2}$.

\smallskip

The equations of motion for the vector perturbations in this case 
 simplify to

\bea\label{pertsdilaton}
&& \frac{(am)^2}{f}\delta \mathcal{A}_0 - \nabla ^2 \delta \mathcal{A}_0 + \boldsymbol{\nabla \cdot} \delta \boldsymbol{\dot{\mathcal{A}}} = 0\,, \\
&& \frac{m^2}{f}\delta \boldsymbol{\mathcal{A}}+ H \delta \boldsymbol{\mathcal{\dot{A}}} + \delta \boldsymbol{\mathcal{\ddot{A}}} + \left(\delta \boldsymbol{\mathcal{\dot{A}}}-\boldsymbol{\nabla}\delta \mathcal{A}_0\right)\frac{\dot{f}}{f}-a^{-2}\nabla^2 \delta \boldsymbol{\mathcal{A}} + 2\left(H+ \frac{\dot{m}}{m}\right)\boldsymbol{\nabla} \delta \mathcal{A}_0 = 0\,. \nn \\
\eea

Moving to Fourier space using Eq.~(\ref{fourier}) as before, the equations for the perturbations (\ref{pertsdilaton}) become
\bea\label{pertsfourierdilaton}
&& \delta{\mathcal{A}}_0 + \frac{i\partial_{t}(\boldsymbol{k \cdot} \delta{\boldsymbol{\mathcal{A}}})}{\left[k^2 + \frac{(am)^2}{f}\right]}=0\,,\\\nn\\
&& \frac{m^2}{f}\delta{\boldsymbol{\mathcal{A}}} + \left(H + \frac{\dot{f}}{f}\right)\delta{\boldsymbol{\mathcal{\dot{A}}}} + \delta{\boldsymbol{\mathcal{\ddot{A}}}}+ \left(\frac{k}{a}\right)^2 \delta{\boldsymbol{\mathcal{A}}}+ \left(2H +2\frac{\dot{m}}{m} - \frac{\dot{f}}{f}\right)\frac{\partial_{t}(\boldsymbol{k \cdot} \delta{\boldsymbol{\mathcal{A}}})}{\left[k^2 + \frac{(am)^2}{f}\right]}=0\,. \nn\\
\eea

We can now compute the equations for the longitudinal and transverse components as before. Using the relations:
\be
\delta{\boldsymbol{\mathcal{A}}}^{||} \equiv \frac{\boldsymbol{k}(\boldsymbol{k \cdot}\delta{\boldsymbol{\mathcal{A}}})}{k^2},
\hspace{1cm}
\delta{\boldsymbol{\mathcal{A}}}^{\bot} \equiv \delta{\boldsymbol{\mathcal{A}}} - \delta{\boldsymbol{\mathcal{A}}}^{||}
\ee
the equations become:
\bea
\label{AVectorperturbs}
&& \left[\frac{m^2}{f} + \left(H + \frac{\dot{f}}{f}\right)\partial_t + \partial_{t}^2 + \left(\frac{k}{a}\right)^2\right]\delta{\boldsymbol{\mathcal{A}}}^{\bot} =0 \\  \nn \\
&&
\left[\frac{m^2}{f} + \left(H + \frac{\dot{f}}{f}\right)\partial_t +\left(2H + 2\frac{\dot{m}}{m}-\frac{\dot{f}}{f}\right)\frac{\left(\frac{k}{a}\right)^2\partial_t}{\left(\frac{k}{a}\right)^2 + \frac{m^2}{f}}+ \partial_{t}^2 + \left(\frac{k}{a}\right)^2\right]\delta{\boldsymbol{\mathcal{A}}}^{||} =0\,, \nn \\
\eea
which are identical to those obtained in Ref.~\cite{Dim1.1}.

At this point we define the spatial components of the canonically normalised, physical (as opposed to comoving) vector field as follows:
\be
\boldsymbol{W} \equiv \sqrt{f}\bcA /a \,.
\label{WA}
\ee
Moving to Fourier space, we expand the perturbations of the physical vector field, $\delta \boldsymbol{W}$, as
 \be
\delta \boldsymbol{W} (t,\boldsymbol{x})=\int \frac{d^3k}{(2 \pi)^{3/2}}\delta \boldsymbol{\mathcal{W}}(t,\boldsymbol{k})e^{i\boldsymbol{k}\cdot \boldsymbol{x}} \,.
\ee
The field equations for the spatial vector perturbations (\ref{AVectorperturbs}) then become
\be
\label{WVectorperturbs}
\left[\partial_t^2 + 3H\partial_t + M^2 + \left(\frac{k}{a}\right)^2\right]\delta{\boldsymbol{\mathcal{W}}}^\perp=0 \,,
\ee
and
\bea
\label{WVectorperturbs+}
\left\{\partial_t^2 + \left[3H + 2H\frac{(\frac{k}{a})^2}{\left(\frac{k}{a}\right)^2 + M^2}\right]\partial_t + M^2 + \left(\frac{k}{a}\right)^2\right\}\delta{\boldsymbol{\mathcal{W}}}^\|=0\,,
\eea
which are the same as those which were found in Ref.~\cite{Dim2}.
Thus in what follows we apply the results in \cite{Dim2} to our present set up.

\section{Statistical Anisotropy}

We have now all the necessary ingredients to study possible cosmological implications of the D-brane set up with varying dilaton discussed in the previous section.
We start by reviewing the relevant results in \cite{Dim2}, which we then apply to our scenario.

Historically, statistical anisotropy due to the contribution of vector field
perturbations to the curvature perturbation was first studied in the context of
the inhomogeneous end of inflation mechanism \cite{yokosoda} (see also
Ref.~\cite{yokosoda+}). However, the first comprehensive mechanism
independent study of the effect of vector field perturbations onto $\zeta$ and
the resulting statistical anisotropy in the spectrum and bispectrum is
presented in Ref.~\cite{stanis}. In the present work, we focus on the vector
curvaton mechanism, which has the advantage of not being restrictive on the
inflation model as well as keeping the inflaton and vector curvaton sectors
independent (they can correspond to two different branes for example)%
\footnote{Indirectly, statistical anisotropy in $\zeta$
can also be generated by considering a mild anisotropisation of the
inflationary expansion, due to the presence of a vector boson field condensate.
In this case, it is the perturbations of the inflaton scalar field which are
rendered statistically anisotropic \cite{anisinf} (for a recent review see 
Ref.~\cite{sodarev}).}.

\subsection{Particle production}

Gravitational particle production of a vector field proceeds analogously to that of a scalar field: vacuum fluctuations are exponentially stretched by the expansion and approach classicality on superhorizon scales, imprinting their spectrum on the homogeneous background. As a realisation of Hawking radiation in a de Sitter background, it can be shown that this process gives rise to the appearance of real particles, which are then interpreted as having been created by the changing gravitational field \cite{hawking}.

In our scenario studying particle production for the vector boson field amounts to
solving Eqs.~(\ref{WVectorperturbs}) and (\ref{WVectorperturbs+}) for the mode
functions of the vector field perturbations. This has been done in
Ref.~\cite{Dim2}, where the following power spectra were obtained:
\be
\mathcal{P}_{L,R}=\left(\frac{H}{2\pi}\right)^2.
\ee
for the transverse components and
\be
\mathcal{P}_{||}=9\left(\frac{H}{M}\right)^2\left(\frac{H}{2 \pi}\right)^2.
\label{Plong}
\ee
for the longitudinal component.
In our present case  $M$ is constant and is
given by Eq.~(\ref{physmass}). It is evident that both power spectra above are
scale-invariant. Comparing the results for the power spectra of
the transverse and longitudinal mode functions, we see that, for $M\ll H$,
\be
\mathcal{P}_{||}\gg \mathcal{P}_{L,R}\;.
\ee
The above results are central to the claim that a vector field modulated by an evolving scalar field can contribute measurable statistical anisotropy to the curvature perturbation.
As we discussed in the previous section, we consider the dilaton as the field which plays the role of a modulating field.

\subsection{The power spectrum}

We assume that the curvature perturbation $\zeta$ receives contributions from both the scalar inflaton field as well as the vector field. In terms of its isotropic and anisotropic contributions, the power spectrum of $\zeta$ may be written as \cite{ack,stanis}
\be
\mathcal{P}_{\zeta}(\boldsymbol{k})=\mathcal{P}_{\zeta}^{\rm iso}(k)
[1 + g(\hat{\boldsymbol{N}}_A \cdot \hat{\boldsymbol{k}})^2]\,,
\ee
where $\mathcal{P}_{\zeta}^{\rm iso}(k)$ is the dominant isotropic contribution, $\hat{\boldsymbol{k}}\equiv \boldsymbol{k}/k$ and $\hat{\boldsymbol{N}}_A \equiv \boldsymbol{N}_A/N_A$ are unit vectors in the directions $\boldsymbol{k}$ and $\boldsymbol{N}_A$ respectively, and $g$ quantifies the degree of statistical anisotropy in the spectrum. The components of $\boldsymbol{N}_A$ are given by $N_A^{i}\equiv \partial N / \partial W_i$, where $W_i$ are the spatial components of the physical vector field (cf. Eq.~(\ref{WA})), $N_A \equiv |\boldsymbol{N}_A|$ and $N$ gives the number of the remaining e-foldings of inflation. The degree of statistical anisotropy in the spectrum may be defined in terms of the various power spectra that contribute to $\mathcal{P}_{\zeta}$ as follows \cite{stanis}
\be
g \equiv N_A^2\frac{\mathcal{P}_{||}-\mathcal{P}_{L,R}}{\mathcal{P}_{\zeta}^{\rm iso}}\,.
\label{g}
\ee
It is clear that, for the case at hand in which $\mathcal{P}_{||}\gg \mathcal{P}_{L,R}$, the degree of statistical anisotropy can be non-negligible. The CMB data provide no more than a weak upper bound on $g$, which allows statistical anisotropy as much as 30\% \cite{GE}. The forthcoming observations of the Planck satellite will reduce this bound down to 2\% if statistical anisotropy is not observed \cite{planck}. This means that, for statistical anisotropy to be observable
in the near future, $g$ must lie in the range
\be
0.02\leq g\leq 0.3\,.
\label{grange}
\ee
Thus, the spectrum is predominantly isotropic but not to a large degree. To avoid generating an amount of statistical anisotropy that is inconsistent with the observational bounds, we then require that the dominant contribution to the curvature perturbation comes from the scalar inflaton.

The vector curvaton contribution to the curvature perturbation is \cite{stanis,VCrev}
\be
\zeta_A=\frac23\hat\Omega_A\frac{W_i\delta W_i}{W^2}\,,
\ee
where \mbox{$\hat\Omega_A\equiv\frac{3\Omega_A}{4-\Omega_A}\simeq\Omega_A$},
with \mbox{$\Omega_A\equiv\rho_A/\rho$} being the density parameter of the
vector field, where $\rho_A$ and $\rho$ are the densities of the vector field
and of the Universe respectively. Therefore, in this case, we should have
\mbox{$\Omega_A\ll 1$} and the contribution of the vector field to $\zeta$ only
serves to imprint statistical anisotropy at a measurable level.

Let us attempt to quantify this contribution in the case of our D-brane vector curvaton
scenario. In Ref.~\cite{stanis} (see also Ref.~\cite{VCrev}) it was shown that,
for the vector curvaton, when \mbox{$\Omega_A\ll 1$} we have
\be
N_A\simeq\frac12\frac{\Omega_A}{W}\,,
\ee
where \mbox{$W=|\mbox{\boldmath $W$}|$} is the modulus of the physical vector
field. In view of the above, Eq.~(\ref{g}) can be recast as
\be
\sqrt g\sim\frac{\Omega_A}{\zeta}\frac{\delta W}{W}\,,
\label{gstan}
\ee
where we have used that \mbox{${\cal P}_\zeta^{\rm iso}\approx\zeta^2$}, since the
curvature perturbation is predominantly isotropic. We also used that
\be
\delta W\sim\sqrt{{\cal P}_\|}\sim H_*^2/M\,,
\label{dW}
\ee
since \mbox{${\cal P}_\|\gg{\cal P}_{L,R}$} and we have employed Eq.~(\ref{Plong}),
with $H_*$ denoting the Hubble scale during inflation.

The contribution of the vector field to $\zeta$ is finalised at the time of
decay of the vector curvaton, since until then it is evolving with time. Thus,
we need to evaluate the above at the time of the vector curvaton decay, which
we denote by `dec'. In Ref.~\cite{Dim2} it was shown that
\be
\Omega_A^{\rm dec}\sim\Omega_A^{\rm end}
\left(\frac{\Gamma}{\Gamma_A}\right)^{1/2}
\min\left\{1,\frac{M}{H_*}\right\}^{2/3}
\min\left\{1,\frac{M}{\Gamma}\right\}^{-1/6},
\label{WAdec}
\ee
where $\Gamma$ and $\Gamma_A$ denote the decay rates of the inflaton and the
vector curvaton fields respectively, and `end' denotes the end of inflation.

During inflation, as long as the dilaton is varying and \mbox{$f\propto a^2$}
and \mbox{$m\propto a$}, the vector curvaton remains frozen with
\be
W=W_*=\sqrt f{\cal A}/a\propto{\cal A}\simeq{\rm constant}\,,
\ee
where we  used Eq.~(\ref{WA}). However, the dilaton is not expected to roll throughout the remaining 50 or so e-foldings of inflation, after the
cosmological scales exit the horizon. Instead, being a spectator field, it
will most probably stop rolling $N_{\rm x}$ e-foldings before the end of
inflation. After this the modulation of $f$ and $m$ ceases and we have
\mbox{$f\rightarrow 1$}. While the mass of the physical vector curvaton
\mbox{$M=m/\sqrt f$} remains constant, this is not so for $W$. Indeed, taking
\mbox{$f=1$}, it is easy to see that $\cal A$ remains frozen. Thus, in view of
Eq.~(\ref{WA}), we find
\be
W\propto 1/a\,.
\ee
It is clear that the same is also true of the vector field perturbation,
{\em i.e.} \mbox{$\delta W\propto 1/a$}. Putting the above together we can estimate
the value of $\Omega_A^{\rm end}$ as follows:
\be
\Omega_A^{\rm end}\sim e^{-2N_{\rm x}}
\left(\frac{M}{H_*}\right)^2
\left(\frac{W_*}{M_P}\right)^2,
\label{WAend}
\ee
where we used the Friedman equation \mbox{$\rho=3(H_*M_P)^2$} and also
that, while the dilaton is varying, the density of the frozen
vector curvaton remains constant with \mbox{$\rho_A\sim M^2W_*^2$} \cite{Dim2}.

Combining the above with Eqs.~(\ref{gstan}), (\ref{dW}) and (\ref{WAdec}) we
obtain
\be
\sqrt g\sim\zeta^{-1}e^{-2N_{\rm x}}
\frac{H_*W_*}{M_P^2}\left(\frac{M}{H_*}\right)^{5/3}
\min\left\{1,\frac{M}{\Gamma}\right\}^{-1/6}
\left(\frac{\Gamma}{\Gamma_A}\right)^{1/2},
\label{g1}
\ee
where `*' denotes the epoch when the cosmological scales leave the horizon and
we considered that \mbox{$M\ll H_*$} for particle production of the vector
curvaton to take place.

\subsection{The bispectrum}

Now, let us consider the bispectrum. As is well known, the bispectrum of the
curvature perturbation is a measure of the non-Gaussianity in $\zeta$ since it
is exactly zero for Gaussian curvature perturbation. This non-Gaussianity is
quantified by the so-called non-linearity parameter $f_{\rm NL}$, which
connects the bispectrum with the power spectra. When we have a contribution of
a vector field to $\zeta$, $f_{\rm NL}$ can be statistically anisotropic
\cite{stanis,fnlanis}. In the case of the vector curvaton with a varying
kinetic function and mass, as is the case in our scenario, it was shown in Ref.~\cite{Dim2}
that non-Gaussianity is predominantly anisotropic. This means that, if
non-Gaussianity is indeed observed without a strong angular modulation on the
microwave sky, scenarios of the present type will be excluded from explaining the dominant contribution
to the non-Gaussian signal.

The value of $f_{\rm NL}$ depends on the configuration of the three momentum
vectors which are used to define the bispectrum. In Ref.~\cite{Dim2} it was
demonstrated that statistical anisotropy is strongest in the so-called
equilateral configuration, where the three momentum vectors are of equal
magnitude. In this case,
\be
\frac65f_{\rm NL}^{\rm eq}=\frac{2g^2}{\Omega_A^{\rm dec}}
\left(\frac{M}{3H_*}\right)^4
\left[1+\frac18\left(\frac{3H_*}{M}\right)^4\hat W_\perp^2\right],
\ee
where $\hat W_\perp$ is the modulus of the projection of the unit vector
\mbox{\mbox{\boldmath $\hat W$}$\,\equiv\,$\mbox{\boldmath $W$}$/W$} onto
the plane determined by the three momentum vectors which define the bispectrum.
From the above, we see that the amplitude of the modulated
$f_{\rm NL}^{\rm eq}$ is \cite{Dim2}
\be
\|f_{\rm NL}^{\rm eq}\|=\frac{5}{24}\frac{g^2}{\Omega_A^{\rm dec}}\,,
\label{fNLamp}
\ee
while the degree of statistical anisotropy in non-Gaussianity is
\be
{\cal G}\equiv \frac18\left(\frac{3H_*}{M}\right)^4\gg 1\,,
\ee
which demonstrates that non-Gaussianity is predominantly anisotropic.

Combining Eqs.~(\ref{g1}) and (\ref{fNLamp}) we can eliminate the dependence
on $W_*$ and obtain
\be
\|f_{\rm NL}^{\rm eq}\|\sim\frac{5}{24}g\zeta^{-2}e^{-2N_{\rm x}}
\left(\frac{H_*}{M_P}\right)^2\left(\frac{M}{H_*}\right)^{2/3}
\min\left\{1,\frac{M}{\Gamma}\right\}^{-1/6}
\left(\frac{\Gamma}{\Gamma_A}\right)^{1/2}.
\label{fNL}
\ee

To proceed further we note that
\be
\frac{H_*}{M_P}<\frac{\delta W}{W}<1\,,
\label{range}
\ee
where the upper bound is to ensure that our perturbative approach remains valid
and the lower bound is due to the requirement that \mbox{$\Omega_A<1$}, \textit{i.e.}
the vector field does not dominate the Universe at any stage. Indeed, since the
ratio \mbox{$\delta W/W$} remains constant throughout the evolution of the
vector field, we find
\be
\frac{\delta W}{W}\approx
\left.\frac{\delta W}{W}\right|_*\sim\frac{H_*^2}{MW_*}\sim
\frac{H_*}{M_P}\frac{1}{\sqrt{\Omega_{A*}}}\,,
\ee
where \mbox{$\Omega_{A*}=(\rho_A/\rho)_*$}. Employing Eq.~(\ref{range}),
Eq.~(\ref{g1}) gives
\be
\!\;
g\zeta^2e^{4N_{\rm x}}\left(\frac{M_P}{H_*}\right)^2\frac{\Gamma_A}{\Gamma}
<\left(\frac{M}{H_*}\right)^{4/3}
\!\!\min\left\{1,\frac{M}{\Gamma}\right\}^{-1/3}\!\!\!<
g\zeta^2e^{4N_{\rm x}}\left(\frac{M_P}{H_*}\right)^4\frac{\Gamma_A}{\Gamma}\,.
\hspace{-1cm}
\label{Mrange}
\ee
Using this, Eq.~(\ref{fNL}) gives
\be
\frac{5}{24}\frac{g^{3/2}}{\zeta}\frac{H_*}{M_P}
<\|f_{\rm NL}^{\rm eq}\|<
\frac{5}{24}\frac{g^{3/2}}{\zeta}\,.
\label{fNLrange}
\ee
From Eq.~(\ref{grange}) the above suggests that
\be
12\leq\|f_{\rm NL}^{\rm eq}\|_{\rm max}\leq 713\,,
\ee
where \mbox{$\zeta=4.8\times 10^{-5}$} is the observed curvature perturbation.
Thus, we see that, if the generated statistical anisotropy in the spectrum is
observable then non-Gaussianity has also a good chance of being observable,
especially since the upper bound in the above is already excessive and violates
the observational constraints, \mbox{$-214<f_{NL}^{\rm eq}<266$} \cite{wmap}.

\subsection{After inflation and reheating}

To explore the possible observational consequences of the type of scenarios we are discussing, we need to consider the evolution of the Universe after the end of inflation.
Once the expansion rate has dropped sufficiently such that $H(t)$  becomes of order $M$, the vector field condensate begins quasi-harmonic coherent oscillations and produces curvaton quanta. It has been shown in Refs.~\cite{Dim1,Dim2} that the energy density and average pressure of the field during the oscillations scale as $\rho_{A}\propto a^{-3}$ and $\overline{p}\approx 0$ respectively, $\ie$ the field behaves as pressureless isotropic matter. After inflation  $f = 1$ and $m = M$. %
Therefore the action for the vector field which is minimally coupled to gravity becomes
\be
S= M_P^{2}\int d^4 x \sqrt{-g}\left(\frac{1}{2}R - \frac{1}{4M_P^2}\cF_{\mu\nu}\cF^{\mu\nu}-\frac{M^2}{2M_P^2}\cA_{\mu}\cA^{\mu}\right)\,.
\ee
We can make a lower bound estimate on the decay rate of the curvaton
field quanta based on gravitational decay, for which the decay rate is given by
\be
\Gamma_{A} \sim \frac{M^3}{M_P^2}\,.
\label{GA}
\ee
The decay products of the vector curvaton are much lighter degrees of freedom
which are, therefore, relativistic.

The physical mass of the vector field in terms of the Planck mass is given in Eq.~(\ref{physmass}), which we quote here:
\be
M = \frac{(2\pi)^5\sqrt{\pi}}{\mathcal{V}_6}M_P\;.
\ee
This should be compared to the inflationary Hubble scale
\be
H=\frac{1}{M_P}\sqrt{\frac{V_0}{3}}\,,
\ee
where $V_0^{1/4}$ %
is the energy scale of inflation.
Therefore, in order to obtain $M \ll H $ we require
\be\label{bound1}
V_0^{1/2}\gg \frac{(2\pi)^5\sqrt{3\pi}}{\mathcal{V}_6}M_P^2\,.
\ee
We can estimate the scale of inflation for slow roll and DBI scalar inflation in order to obtain a bound on the compact volume.

$\bullet$ In a slow roll inflationary scenario, the CMB observations suggest
\cite{book}
\be
V_0^{1/4}=0.027\epsilon^{1/4}M_P\;,
\label{infscale}
\ee
where $\epsilon$ is the slow-roll parameter. Assuming that $\epsilon$ is not tiny ({\it e.g.}~taking \mbox{$\epsilon>10^{-5}$}) we obtain \mbox{$V_0^{1/4}\sim 10^{16}\,$GeV}. Thus, for \mbox{$M\ll H$} to be fulfilled, we require that the size of the dimensionless volume $\mathcal{V}_6$ is around $10^8-10^9$ or larger, in which case the physical mass of the vector field is $M<10^{14}\rm{GeV}$.
For such a physical mass, Eq.~(\ref{GA}) suggests
\mbox{$\Gamma_{A}<10^6 \rm{GeV}$}. The temperature at the time of
the vector curvaton decay is \mbox{$T_{\rm dec}\sim\sqrt{M_P\Gamma_A}$}. It is
easy to see that the decay occurs before Big Bang Nucleosynthesis if
\mbox{$M> 10\,$TeV}.

$\bullet$ If inflation is instead driven by a DBI scenario, an equivalent expression to (\ref{infscale}) can be found (see for example \cite{chks})  to be
\be
V_0^{1/2}=0.03 \, \tilde h^{-1/4} M_P^2\;,
\label{dbiscale}
\ee
where $\tilde h = h \,M_P^4$ is a dimensionless warp factor. For a GUT inflationary scale, which is consistent with DBI inflation, we require $\tilde h^{1/4} \sim 10^3$, which can be achieved near the tip of the throat, where the warp factor is larger. Using this, we again obtain a limit on the 6D compact volume of the same order as above, and thus analogous results follow.
Moreover, if inflation happens in a DBI fashion, a second source for large non-Gaussianites of equilateral shape is generated \cite{st}.

\bigskip

What about the inflaton decay, which reheats the Universe (since the vector curvaton is always subdominant)? In the case where the vector field brane is
static, inflation is driven by some other sector. Possibly, this is another
D-brane undergoing motion along the warped throat that may be either of the slow roll or DBI variety.
The end of inflation takes place when the inflaton brane approaches the tip of the throat, where there is an IR cutoff
which allows the brane to reach the origin at finite time and oscillate around the minimum of the potential \cite{st,muko,EGMTZ}. In this case, the inflaton decay rate
$\Gamma$ depends on the couplings of the inflaton to standard model particles, to which it decays. A more dramatic end of inflation can
take place if the inflaton brane meets and annihilates with the antibrane located at the end of the tip. This is a complex process which occurs via a cascade that begins with a gas of closed strings, followed by Kaluza-Klein modes and eventually standard model particles \cite{Dreheating}.

The above possibilities may also arise in the case that our vector curvaton field is living on a moving brane, which can indeed be the inflaton brane, as will be subsequently discussed. However, in this case, prompt reheating by annihilation is not possible because we need the vector field to survive after the end of inflation in order to play the role of the curvaton.

\subsection{Examples}

We are now ready to illustrate through a few examples how the D-brane curvaton toy model we presented in section III.B can lead to observable statistical anisotropy in the spectrum and bispectrum of the
curvature perturbation. In all cases we consider inflation at the scale of
grand unification, with \mbox{$H_*\sim 10^{14}\,$GeV}, which is favoured by
observations [cf. Eq.~(\ref{infscale})].

Even though there is no compelling reason why the value
of the physical vector field cannot be super-Planckian
(since \mbox{$\Omega_A<1$})\footnote{This is in contrast to scalar fields,
where super-Planckian values are expected to blow-up non-renormalisable terms
in the scalar potential and render the perturbative approach invalid.}, we make
the conservative choice \mbox{$W_*\lesssim M_P$} in the examples below. %
As mentioned above, the dilaton is not expected to continue to roll for the remaining 50 or so e-folds after the cosmological scales exit the horizon. The cosmological scales span about 10 e-folds. Therefore, for the signal to be generated and span all the cosmological scales, one would strictly require that the dilaton rolls appropriately for up to 10 e-folds.\footnote{Note, however, that it is possible for statistical anisotropy not to span all of the cosmological scales. For example, it might be there only for very large scales, comparable to the present horizon.}
 However,  if inflation is continuous,  $N_{\rm x}$ is expected to be small, and the dilaton will need to roll for a substantial number of e-folds to allow  for  observable statistical anisotropy. 
 On the other hand, if there is an early phase of inflation that lasts for up to 10 e-folds after which the vector field decays and imprints the spectrum, then the modulation period needs only to last for this early phase, for once the spectrum is imprinted it can no longer be diluted by further bouts of inflation. While there are several motivations in string theory for considering successive periods of inflation, such as bouncing branes or inflation from a subset of the appropriately light moduli, in what follows we will take the simplest case and assume that inflation is continuous, and will briefly comment at the end on cases where inflation is not continuous.

\subsubsection{Prompt reheating}

Let us consider first the case of prompt reheating, where we simply assume that all of the initial vacuum energy of the branes is converted into
radiation \cite{earlyReheating}.  This would lead to almost instantaneous reheating (prompt reheating) with a reheating temperature
 \mbox{$T_{\rm reh}\sim V_0^{1/4}$}, which implies \mbox{$\Gamma\sim H_*$}.

Then, using also Eq.~(\ref{GA}), we can recast Eq.~(\ref{Mrange}) as
\be
\frac{e^{-2N_{\rm x}}}{\sqrt{g\zeta^2}}\frac{H_*^2}{M_P}<M<
\frac{e^{-2N_{\rm x}}}{\sqrt{g\zeta^2}}H_*\;.
\ee
Using that \mbox{10 TeV$\,<M<H_*$}, the above gives
\be
\ln\left[(g\zeta^2)^{-1/4}\sqrt{\frac{H_*}{M_P}}\right]\lsim N_{\rm x}\lsim
\ln\left[(g\zeta^2)^{-1/4}\sqrt{\frac{H_*}{10\,{\rm TeV}}}\right].
\ee
Employing Eq.~(\ref{grange}) and considering \mbox{$H_*\sim 10^{14}\,$GeV},
we obtain
\be
1\lsim N_{\rm x}\lsim 18\,.
\ee
This range is further truncated if we postulate \mbox{$W_*\lsim M_P$} as
mentioned. Now, with prompt reheating Eq.~(\ref{g1}) becomes
\be
g\sim\zeta^{-2}e^{-4N_{\rm x}}\left(\frac{W_*}{M_P}\right)^2,
\ee
which is independent of the value of $M$.
If we choose \mbox{$W_*\sim M_P$} then \mbox{$g\sim 0.1$} for
\mbox{$N_{\rm x}\approx 6$}. With these values Eq.~(\ref{fNL}) gives
\be
\|f_{\rm NL}^{\rm eq}\|\sim 0.01\times\left(\frac{H_*}{M}\right).
\ee
Thus, we can obtain observable non-Gaussianity ($\|f_{\rm NL}^{\rm eq}\| \gtrsim 1$)  for
\mbox{$10^{10}\,$GeV$\,\leq M\leq 10^{12}\,$GeV}.

\subsubsection{Intermediate reheating scale}

Let us consider another example, where  we now assume \mbox{$\Gamma\sim M$}. Then, following the same
process as above, we arrive at
\be
\ln\left[(g\zeta^2)^{-1/4}\sqrt{\frac{H_*}{M_P}}\right]\lsim N_{\rm x}\lsim
\ln\left[(g\zeta^2)^{-1/4}\left(\frac{H_*}{10\,{\rm TeV}}\right)^{1/6}\right].
\ee
Employing Eq.~(\ref{grange}) and considering \mbox{$H_*\sim 10^{14}\,$GeV},
we obtain
\be
1\lsim N_{\rm x}\lsim 10\,.
\ee
This range is further truncated if we postulate \mbox{$W_*\lsim M_P$}.

With \mbox{$W_*\sim M_P$}, Eq.~(\ref{g1}) becomes
\be
g\sim\zeta^{-2}e^{-4N_{\rm x}}\left(\frac{M}{H_*}\right)^{4/3}.
\label{g2}
\ee
Combining this with Eq.~(\ref{fNL}) we obtain
\be
\|f_{\rm NL}^{\rm eq}\|\sim\frac{5}{24}\zeta^{-4}e^{-6N_{\rm x}}
\left(\frac{M}{M_P}\right).
\ee
Using \mbox{$N_{\rm x}\approx 4$}, we find that
\mbox{$\|f_{\rm NL}^{\rm eq}\|\sim 100$} can be attained if
\mbox{$M\sim 10^{-7}M_P\sim 10^{11}\,$GeV}.
Using this value, Eq.~(\ref{g2})
gives \mbox{$g\sim 0.02$}.
These are just sample ``large" values,
they can become smaller if $M$ is reduced.

\subsubsection{Gravitational reheating}

As a last example, we assume now that the inflaton decays through gravitational couplings. If we
further take the inflaton mass to be of order $H_*$ (this is natural in
supergravity \cite{randall}), then we have
\be
\Gamma\sim\frac{H_*^3}{M_P^2}\,.
\label{Ginf}
\end{equation}
For inflation at the scale of grand unification we find
\mbox{$\Gamma\sim 10^6\,$GeV}.

Now, let us consider that the dilaton rolls throughout inflation, {\it i.e.}~\mbox{$N_{\rm x}=0$}. Taking \mbox{$W_*\sim M_P$}, Eq.~(\ref{g1}) becomes
\be
\frac{M}{H_*}\sim(g\zeta^2)^3\left(\frac{M_P}{H_*}\right)^6,
\ee
where we have used Eq.~(\ref{GA}) and we have assumed \mbox{$M>\Gamma$}.
Employing Eq.~(\ref{grange}) and considering \mbox{$H_*\sim 10^{14}\,$GeV}, the
above gives
\be
10^{-7}H_*\lesssim M\lesssim 10^{-4}H_*\;.
\ee
for an observable signal.
Let us take \mbox{$M\sim 10^{-4}H_*\sim 10^{10}\,$GeV$\,>\Gamma$}, which means
\mbox{$\Gamma_A\sim 10^{-6}\,$GeV}, according to Eq.~(\ref{GA}). Then,
Eq.~(\ref{g1}) yields \mbox{$g\sim 0.1$}, while Eq.~(\ref{fNL}) gives
\mbox{$\|f_{\rm NL}^{\rm eq}\|\sim 10^2$}. Again, these values
can be reduced for a smaller $M$.

From the above examples we see that it is  possible to generate
observable statistical anisotropy in the spectrum and bispectrum of the
curvature perturbation. If we allow for super-Planckian $W_*$ we can increase
$N_{\rm x}$ but no more than \mbox{$N_{\rm x}\simeq 20$}.
Therefore
the dilaton needs to roll for a substantial number of e-foldings to allow for
observable statistical anisotropy. This problem is ameliorated if there is a
subsequent period of inflation (\textit{e.g.} thermal inflation, which can
contribute about 20 e-foldings or so \cite{thermal}). In particular, for
super-Planckian $W_*$ and for a total period of inflation that lasts
$N \gtrsim60$ e-foldings, of which 20 e-foldings may be generated subsequently
by thermal inflation for example, we see that the dilaton must evolve for
about 20 e-foldings.
Further improvement can be achieved if one
considers several bouts of inflation, {\em e.g.}~as in \cite{EGMTZ}. The reheating temperature
\mbox{$T_{\rm reh}\sim\sqrt{M_P\Gamma}$} in the above examples is rather large
and would result in an overproduction of gravitinos, if the latter were stable.
This problem is also overcome by adding a late period of inflation since the
entropy release can dilute the gravitinos.

\section{Moving brane}

We now consider the brane whose world volume hosts our vector curvaton to be the D3-brane which is driving inflation. In open string D-brane models of
inflation, inflationary trajectories can arise from motion in the radial direction of a warped throat, in which brane motion may be slow or relativistic,
leading to slow-roll or DBI inflation respectively%
\footnote{As previously mentioned, in a realistic scenario where the St\"uckelberg mass mechanism is used, one would need to consider branes of higher dimensionality. For D-brane inflation in a warped throat as we consider  here, one may consider the inflaton brane to be wrapped D5-brane as in Ref.~\cite{becker}.}. In addition to the radial direction, one may also consider the brane to have non-trivial motion in any of the five angular directions of the throat, giving an inherently multifield scenario. As shown in Ref.~\cite{EGMTZ,GK} for the case of radial motion plus one angular field, motion in the angular directions experiences strong Hubble damping such that the behaviour of the brane very soon tends towards the conventional single field scenario. This situation was recently confirmed in Ref.~\cite{McAB} in which motion in all six directions in the throat is considered.
Thus we can assume that for most of the  inflationary period, the motion of the brane in the throat is effectively along a single direction. Nevertheless, it is useful to comment on the multifield case since these scenarios overcome the essential problems with single field DBI inflation, which, for example, are related to the lack of consistency of predicted bounds on the scalar-to-tensor ratio \cite{LH}.
In what follows, we briefly outline the two  possibilities, \textit{i.e.} slow-roll inflation and DBI inflation, where we consider both single field and multifield DBI scenarios. All the results from the previous sections can then just follow straightforwardly.

\smallskip

To take into account the possible effects of relocating our vector curvaton to a moving brane, we consider the same scenario as is discussed in Sec.~\ref{subsec:varying_dilaton} ({\it i.e.}~a canonical vector field modulated by the dilaton field such that $e^{-\phi} \propto a^2$), but with a position field for general brane motion in the radial direction. We therefore rewrite the action in Eq.~(\ref{total_action}) as
\begin{align}
\label{smlF_action}
\begin{split}
 S = &\int \!d^4 x \sqrt{-g} \left\{ \frac{M_P^2}{2}R -\frac{M_P^2}{4}\partial_{\mu}\phi \,\partial^{\mu}\phi - V(\phi)- h^{-1}\left[1 + he^{-\phi}\cF_{\alpha\beta}\cF^{\alpha\beta}+ h\partial_\alpha\varphi\partial^\alpha \varphi \right.\right.
\\
& \left.\left. +  3h^2e^{-\phi}\left(\partial_\alpha \varphi\partial^\alpha \varphi \cF_{\beta\gamma}\cF^{\beta\gamma} - 2\partial_\alpha \varphi \cF^{\alpha\beta}\partial^\gamma \varphi \cF_{\gamma\beta}\right)\right]^{1/2}- V(\varphi)+ h^{-1}  -  \frac{m^2 }{2} {\mathcal A}_\mu  {\mathcal A}^\mu \right\}.
\end{split}
\end{align}
We consider cosmological scenarios such that all background fields are functions of time only, in which case there is a cancelation of the terms mixed in $\varphi$ and $\cA_\mu$ at background level. Nonetheless, in principle this action could still lead to mixed terms in the perturbations, which can be functions of space as well as time. It turns out however, that the mixed terms do not appear in the equations of motion for the perturbations of both $\varphi$ and $\cA_\mu$, as is clear from examining the complete equations of motion which are given below and considering the possible perturbations of the various terms.

The equations of motion for $\varphi$ and $\cA_\mu$ calculated from the action in Eq.~(\ref{smlF_action}) are given respectively by,
\bea
\label{GeneralEOMs}
&&\hskip-0.6cm \frac{h'}{h^2} \!\!\left(1-\sqrt{\Sigma}\right)\!+\!V'(\varphi) +\! \frac{h'}{h}\frac{e^{-\phi}\cF_{\alpha\beta}\cF^{\alpha\beta}+\partial_\alpha \varphi \partial^\alpha \varphi+ 6he^{-\phi}[(\partial_\alpha\varphi)^2 \cF^2 - 2\partial_\alpha \varphi \cF^{\alpha\beta}\partial^{\gamma}\varphi \cF_{\gamma\beta}]}{2\sqrt{\Sigma}}\nn
\\ \nn
\\
&& \hspace{2cm}=  \frac{\partial_{\mu}}{2\sqrt{-g}}\left\{\frac{\sqrt{-g}[2\partial^\mu \varphi + 3he^{-\phi}(2\cF_{\alpha\beta}\cF^{\alpha\beta}\partial^\mu \varphi - 4\partial^\alpha \varphi \cF_{\alpha \beta}\cF^{\mu \beta})]}{\sqrt{\Sigma}}\right\}
\eea
\be
\!\! m^2 \cA^\nu = \frac{\partial_{\mu}}{\sqrt{-g}}\left\{\frac{\sqrt{-g}(e^{-\phi}\cF^{\mu\nu}+3he^{-\phi}[2(\partial_{\alpha}\varphi)^2\cF^{\mu\nu}-2\partial_{\alpha}\varphi \cF^{\alpha \nu}\partial^{\mu}\varphi+2\partial_{\alpha}\varphi \cF^{\alpha \mu}\partial^{\nu}\varphi)]}{\sqrt{\Sigma}}\right\},
\ee
where
\be
\hspace{1cm}\Sigma = 1 + h(\partial_\alpha\varphi)^2 + he^{-\phi} \cF^2+3h^2e^{-\phi}[(\partial_\alpha\varphi)^2 \cF^2 - 2\partial_\alpha \varphi \cF^{\alpha\beta}\partial^{\gamma}\varphi \cF_{\gamma\beta} ]\,.\nn
\ee
Neglecting the mixed terms at both background and perturbation level, and considering the derivatives acting on $\cA_\mu$ to be small while keeping those acting on the position field to be general, we may expand
$\sqrt{\Sigma}$, and, keeping only up to quadratic order in $\cF$, the equations of motion then become
\bea
&&\hskip-0.4cm
V'(\varphi) + \frac{h'}{h^2}\!\left[1\!-\!\sqrt{1+ h(\partial_\alpha \varphi)^2}\right]\!+\! \frac{h'}{2h}\frac{\left(e^{-\phi}\cF_{\alpha\beta}\cF^{\alpha\beta}+ \partial_{\alpha}\varphi \partial^\alpha \varphi\right)}{\sqrt{1+ h(\partial_\alpha \varphi)^2}} =  \frac{\partial_{\mu}}{\sqrt{-g}}\!\left[\!\frac{\sqrt{-g}\partial^\mu \varphi}{\sqrt{1 + h(\partial_\alpha\varphi)^2}}\right],\nn 
\eea
for the inflaton field, and
\be\label{DBIvector}
m^2 \cA^\nu = \frac{\partial_{\mu}}{\sqrt{-g}}\left[\frac{\sqrt{-g}e^{-\phi}\cF^{\mu\nu}}{\sqrt{1 + h(\partial_\alpha\varphi)^2}}\right],
\ee
for the vector field.

An important feature of (\ref{DBIvector}) to note is the new form of the gauge kinetic function
\begin{align}
f = \frac{e^{-\phi}} {\sqrt{1+ h\,\partial_\alpha\varphi\partial^\alpha \varphi}}=
    \gamma_\varphi  \,e^{-\phi}.
\end{align}
Employing the metric (\ref{metrica}) into these equations, we find that the equations of motion for the background fields $\varphi (t)$ and $\cA_\mu (t)$ are given respectively by
\be
\label{backgroundDBI}
\ddot{\varphi}-\frac{h'}{h^2}+\frac{3}{2}\frac{h'}{h}\dot\varphi^2 + 3H\dot\varphi \frac{1}{\gamma_\varphi^2} -\frac{h'}{h}e^{-\phi}\left(\frac{\dbcA}{\gamma_\varphi a}\right)^2 + \left(V'(\varphi)+ \frac{h'}{h^2}\right)\frac{1}{\gamma_\varphi^3}=0\,
\ee
and
\be
\cA_t=0\,,
\ee
\be
\label{backgroundADBI}
\ddbcA + \dbcA\left(H + \frac{\dot f}{f} \right) + \frac{m^2}{f}\bcA =0\,.
\ee
In the absence of the vector field, Eq.~(\ref{backgroundDBI}) reduces to the standard equation of motion for the DBI inflaton (see Ref.~\cite{st}),
where we note that now, since the scalar field is homogenised by inflation,
\be\label{DBIgamma}
\gamma_{\varphi} = \frac{1}{\sqrt{1 - h \,\dot{\varphi}^2}}\,.
\ee
Eq.~(\ref{DBIgamma})  gives the Lorentz factor for brane motion in the internal space, and is a direct generalisation of the Lorentz factor for a relativistic point particle. In an AdS throat, the warp factor is simply given by
\be
\label{warpAdS}
h = \frac{\lambda}{\varphi^4}
\ee
where $\lambda=g^2_{YM}$ is the 't Hooft coupling, and we require $\lambda \gg 1$ such that the system may be described via the gravity side of the AdS/CFT correspondence. The brane must obey the causal speed limit in the bulk, which is equivalent to requirement that $\gamma_\varphi$ remains real at all times. Given the form of the warp factor in Eq.~(\ref{warpAdS}), we see that warping becomes significant as $\varphi \rightarrow 0$, therefore at small $\varphi$ the velocity of the relativistic brane is forced to decrease. Indeed, in Ref.~\cite{st} it is shown that at late times
\be
\varphi(t)\rightarrow \frac{1}{t},
\ee
which implies that for a pure AdS throat, the brane takes an infinite time to cross the horizon. A realistic throat may be approximated as AdS in the regions of interest but has a finite cut-off at the IR end, therefore the brane may cross the horizon in finite time. The fact that the brane is forced to slow down as it moves towards the horizon leads to inflationary trajectories in this region.

In an AdS geometry, $\gamma_\varphi$ may indeed become arbitrarily large at late times, and this leads to a suppression of all but the first three terms in Eq.~(\ref{backgroundDBI}). Therefore, we see that in this case the vector term has a negligible impact on the dynamics of the inflaton, along with the potential term and friction term, as soon as the brane starts to approach the speed limit $\dot{\varphi}=\varphi^2/\sqrt{\lambda}$. The same is true for the potential and friction terms in the case of standard DBI inflation in an AdS throat (see Ref.~\cite{st}).

In a Klebanov-Strassler throat, the behaviour of $\gamma_\varphi$ is such that its maximum value is reached almost immediately as the brane moves from the UV end of the throat, dropping for subsequent times. At late times, when the brane is moving in the IR region of the throat, the value of $\gamma_\varphi$ is roughly constant, remaining within a single order of magnitude. Ultimately $\gamma_\varphi \rightarrow 1$, as the brane stops. This means that the vector, friction and potential terms in Eq.~(\ref{backgroundDBI}) are no longer suppressed for later times. In this case the vector $\mathcal{A}_\mu$ can have an influence on the dynamics of the inflaton.
In the standard DBI scenarios in Klebanov-Strassler throats, \textit{i.e.} without the vector contribution, the presence of non-negligible potential and friction terms in the dynamics of the inflaton does not change the result: inflation still takes place in the throat. For our case, the presence of the vector field may contribute an effective term in the potential for the inflaton, along the lines of what has been demonstrated in Ref.~\cite{Dim3}. For the time being we focus on the simpler AdS case, such that the vector term is subdominant in the dynamics of the inflaton, and we can treat the system as undergoing standard DBI inflation in an AdS background. However, further work is currently in progress that will assess the impact of the vector backreaction in a Klebanov-Strassler throat.

Let us now consider conventional slow-roll, which is possible if the potential admits a particularly flat section. When the brane is slowly rolling along a flat section of its potential in the throat,  the derivatives of both the vector as well as the position field are small  and we can expand the $\sqrt{\Sigma}$ factor in Eqs.~(\ref{GeneralEOMs}). Keeping only those terms that are up to quadratic order in the derivatives of both of the fields,
 we  recover the standard Klein-Gordon equation for a minimally coupled scalar field,
\be
\ddot{\varphi} + 3H\dot\varphi + V'(\varphi)=0\,.
\ee

We can now implement the vector curvaton scenario in this set-up as follows.
Assuming that the vector field and the dilaton give a subdominant contribution to the energy density during inflation, the energy density and pressure calculated from the action in Eq.~(\ref{smlF_action}) are given by
\be\label{energydensity}
\rho = \frac{1}{h}\left(\gamma_{\varphi} - 1\right) + V\,,
\ee
\be\label{pressure}
p = \frac{1}{h}\left(1-\frac{1}{\gamma_{\varphi}}\right) - V\,.
\ee
We consider the brane to be moving relativistically, therefore $\gamma_\varphi$ is large. As discussed above, in the limit of strong warping the velocity of the brane is forced to decrease, hence the energy density becomes dominated by the potential. For large $\gamma_\varphi$ and strong warping, the pressure is clearly also dominated by the potential. This illustrates how inflation can arise in a DBI scenario.

Taking into account the new form of the gauge kinetic function, we see that if $\gamma_{\varphi} \neq 1$ the scaling necessary for statistical anisotropy could in principle be spoilt by new powers of the scale factor that are introduced as a result of the inflaton. As shown in Ref.~\cite{st}, for DBI inflation in an AdS geometry with a warp factor as in Eq.~(\ref{warpAdS}), the scale factor $a(t)\rightarrow a_0 \,t^{1/\epsilon}$ at late times, where $\epsilon$ is a generalisation of the slow-roll parameter and is given by
\be
\epsilon = \frac{2 M_{P}^2}{\gamma_\varphi}\left(\frac{H'}{H}\right)^2,
\ee
such that $\ddot{a}/a=H^2(1-\epsilon)$ and one obtains de Sitter expansion for $\epsilon \rightarrow 0$. For a background expanding in this way, the vector field will undergo gravitational particle production as outlined in Sec.~\ref{subsec:varying_dilaton} and obtain a scale invariant spectrum of superhorizon perturbations as long as we still have $f \propto a^2$ and as long as the vector field remains light (the vector mass $m$ does not depend on the inflaton and therefore the condition $m \propto a$ is not impacted, however the physical mass $M = m/\sqrt{f}$ is impacted).

It is further shown in Ref.~\cite{st} that $\gamma_\varphi \propto t^2$ at late times, \textit{i.e.}~$\gamma_\varphi \propto a^{2\epsilon}$ which means that our gauge kinetic function is now $f= e^{-\phi}\,\gamma_\varphi \propto a^{2+2\epsilon}$. This could contribute a small degree of scale dependance to the power spectrum of vector perturbations, however clearly the scaling $f \propto a^2$ still holds. Furthermore,
as shown in Ref.~\cite{Dim2}, when \mbox{$f\propto a^{2(1+\epsilon)}$}, the spectral tilt for the transverse and longitudinal components of the vector field are different with the corresponding spectral indexes being
\be
n_{L,R}-1=-\frac83\epsilon\quad{\rm and}\quad n_\|-1=2\epsilon\,,
\ee
{\it i.e.}~the transverse spectrum is slightly red and the longitudinal is slightly blue.

 The physical mass of the vector field is now given by $M = m /\sqrt{f} \propto 1/\sqrt{\gamma_\varphi} \propto a^{-\epsilon}$, which means that $M$ now experiences a slight evolution during inflation. In particular, when $\gamma_\varphi \gg 1$ the magnitude of the vector mass is decreased such that any evolution of $M$ only serves to make the condition $M \ll H$ easier to fulfill.

In addition to the suppression of the vector term by $\gamma_\varphi$, we saw before that for a physical vector mass $M \ll H$ during inflation and a gauge kinetic function $f \propto a^{2}$, the equation of motion for $\cA_\mu$, given in Eq.~(\ref{backgroundA}), implies that the vector field freezes at constant amplitude, such that $\dot\cA_\mu$ (which appears in Eq.(\ref{backgroundDBI})) is expected to be very small during this time. The same behaviour occurred during inflation for the non-canonical vector field, as can be seen in Fig.~\ref{fig. 1}. 
Note also, that the vector field is coupled to the inflaton through the Lorentz factor in Eq.~(\ref{DBIgamma}), which features the derivatives of $\varphi$, which are expected to be small during slow-roll inflation.\footnote{This is in contrast to Ref.~\cite{Dim3},
where the kinetic function of the vector field is modulated by the inflaton
field itself and not by its derivatives.}

In Refs.~\cite{BDS,DPotential} the potential for a D3-brane has been explicitly calculated taking into account all corrections from fluxes and bulk objects, and the results show that it is  possible, albeit with fine-tuning, to obtain a flat region in which a slow-roll phase could occur. Similarly, an explicitly calculated potential is studied in Ref.~\cite{EP}, in which it is shown that a sufficiently long period of inflation as well as a correct spectrum of perturbations can be achieved from the combination of a slow-roll and DBI phase, where slow-roll is obtained by fine-tuning the potential in the region close to the tip. In such pictures in which the dominant contribution to the curvature perturbation can be successfully generated by the inflaton,
 modulated vector curvaton
of the type considered in the present work could add the new feature of measurable statistical anisotropy.

\smallskip

Let us now comment on the multifield case.
In the simplest situation one may consider a generalization to a two field model in which, in addition to its motion in the radial direction, the brane moves in one of the five angular directions of the warped throat. Such a scenario is considered in Ref.~\cite{EGMTZ,GK} and we briefly discuss this picture here to illustrate the multifield generalisation of our work, keeping in mind that the dominant behaviour of the brane is always well approximated by a single field scenario at the times of interest to us.

In general for a multifield scenario, several of the terms contained in the determinant in Eq.~(\ref{determinant}) computed for the DBI action in Eq.~(\ref{DBI}) may no longer vanish after the antisymmetrisation.  However, these terms become subdominant as soon as the brane tries to move radially only, and therefore we do not need to consider them to explain the essentials of the multifield picture. The important point is that the vector field and the scalar fields are decoupled for late times at both background and perturbation level, as we saw for the single field case. Considering only the dominant terms in Eq.~(\ref{determinant}), the transition to a multifield scenario will impact the form of $\gamma_{\varphi}$, which now becomes,
\be
\label{gammamulti}
\gamma_{\varphi} = \frac{1}{\sqrt{1-h\, \dot{\varphi}^i \dot{\varphi}^jg_{ij}}}\,,
\ee
where $g_{ij}$ is the metric on the internal space.

The energy density and pressure calculated from this action are analogous to Eqs.~(\ref{energydensity}) and (\ref{pressure}), but where $\gamma_{\varphi}$ is now given by (\ref{gammamulti}).
Once the brane starts to settle onto a radial trajectory, we recover single field DBI inflation and the familiar form of $\gamma_{\varphi}$ given in (\ref{DBIgamma}). All of the results outlined before for the single field case are then applicable here.

\section{Conclusions}

In this paper we have discussed  the possibility to embed the vector curvaton scenario in string theory
where the vector field which  lives
on a D3-brane plays the role of the vector curvaton.
We focused on the simplest case which may outline the mechanism within string theory, and which may be generalised to realistic explicit cases, or used as a means to inform the search for such cases. We have investigated how this scenario can affect the observed curvature
perturbation $\zeta$ in the universe.
We have first considered the case in which the vector curvaton brane is stationary and inflation occurs in some other sector,
for example via warped ${\rm D}\bar{\rm D}$ inflation, or via the motion of a different D3-brane.
For suitable
values of the parameters, such a vector curvaton can generate
observable statistical anisotropy in the spectrum and bispectrum of $\zeta$
provided that the dilaton field, which is a spectator field during inflation, varies
with the scale factor as \mbox{$e^\phi\propto a^{-2}$} when the
cosmological scales exit the horizon. If this is the case,
both the transverse and the longitudinal components of the vector field obtain a scale-invariant superhorizon spectrum of perturbations. However, particle production is anisotropic, which means that the vector curvaton cannot generate $\zeta$ by itself but it can give rise to observable statistical anisotropy. In view of the forthcoming data from the Planck satellite, this is a finding that will be testable in the very near future. Indeed,
in the class of models we have discussed, statistical anisotropy in the bispectrum is predominant. This means
that non-Gaussianity has to have a strong angular modulation on the microwave sky, which may or may not be found by Planck. Moreover,  these models can still produce observable statistical anisotropy in the spectrum even if its
contribution to the bispectrum is negligible (and vice-versa, see also
Ref.~\cite{gvsfNL}).

We also showed that these results are robust when we allow for the possibility that the same brane which hosts the vector curvaton,
is also responsible for driving cosmological inflation, which  can be of either the slow roll or the DBI variety.
All the results obtained for the stationary brane case follow and we again can obtain measurable statistical anisotropy both in the
spectrum and the bispectrum.
Moreover, in the case of DBI inflation, the constraints on the vector mass can be considerably improved. Furthermore, since  DBI inflation
also contributes to the generation of large non-Gaussianities of the equilateral type, in this case there would be two different sources for large Gaussian deviations.
Thus we have seen that the presence of several light fields in string theory models of cosmology can provide us with a unique source for
distinctive features, which can help us to distinguish such models from pure  field theory models. Certainly these possibilities deserve
further investigation, and in the present work we have only begun to explore the prospects for sources of stringy statistical anisotropies
in the curvature perturbation.

As we have shown, our results are  based on the assumption that the dilaton rolls as $e^{-\phi}\propto a^2$
while the cosmological scales exit the horizon, and then for a further case-dependent amount of time %
in order for statistical anisotropy to be observable.
 Nevertheless, for the simplest cases in which inflation is continuous, we require that the minimum of the dilaton potential is located at an extremely  weak coupling, $g_s\ll1$ for an observable signal.
We have commented that this is improved for cases in which
inflation occurs in successive periods, which are not uncommon among string theory models.

We focused on vector fields living on D3-branes, which acquire a mass via a St\"uckelberg mechanism. In this set up, we considered the dilaton as a modulating field for the mass and gauge kinetic function of the vector field. We have not considered the details of compactification and moduli stabilisation, which are clearly necessary for a concrete realisation of our work. Thus as we have already mentioned, the present work should be considered as a first step towards building more explicit models within string theory of the vector curvaton paradigm,
where the viability of such models is explored, and attention is drawn to the possibility that distinctive features might arise in the curvature perturbation as a result of D-brane vector fields. Along these lines, for the vector curvaton paradigm to work, the key requirements are that the vector field is light while the scales exit the horizon, and that its mass and gauge kinetic function are modulated in a non-trivial way by a scalar degree of freedom or modulon, such that the resulting spectrum is scale invariant and thus observationally consistent. In our simple picture, we have shown that it is indeed possible, in principle, to meet all of these requirements for vector fields in D-brane models of inflation in Type IIB string theory, assuming that the dilaton can behave accordingly. The striking feature is that the dilaton appears with just the right powers in the gauge kinetic function and mass of the vector field to make scale invariance possible, thereby making the D-brane vector curvaton a viable prospect. This strongly motivates a further and more concrete exploration of vector curvatons in string theory, and as we have already stressed, the dilaton is merely the most obvious choice for a modulon in a scenario where simplicity is favored over concreteness.

\acknowledgments
We would like to thank M.~Blaszczyk, T.~Koivisto, A.~Maharana, D.~Mayorga-Pe\~na, P.~Oehlmann, F.~Quevedo, F.~Ruehle, M.~Schmitz and G.~Tasinato for helpful discussions.
K.D.~wishes to thank the University of Crete for the hospitality.
K.D~is supported by the Lancaster-Manchester-Sheffield Consortium for
Fundamental Physics under STFC grant ST/J000418/1.
D.W.~and I.Z.~are supported by the DFG cluster of excellence
Origin and Structure of the Universe, the SFB-Tansregio TR33
``The Dark Universe" (Deutsche Forschungsgemeinschaft) and
the European Union 7th network program ``Unification in the
LHC era" (PITN-GA-2009-237920). D.W. would also like to thank BIGS-PA for support.

\appendix
\section{ Mass generation mechanism for $U(1)$ field }
\label{A1}

Consider a Lagrangian of the following form \cite{GIIQ}:
\be\label{mass1}
\mathcal{L}=-\frac{e^{\phi}}{12}H^{\mu\nu\rho}H_{\mu\nu\rho}
-\frac{e^{-\phi}}{4\,\tilde g^2}F^{\mu\nu}F_{\mu\nu}+\frac{c}{4}\epsilon^{\mu\nu\rho\sigma}B_{\mu\nu}F_{\rho\sigma}\;,
\ee
where $F_{\mu\nu}= \partial_{\mu}A_{\nu}- \partial_{\mu}A_{\nu}$, $H_{\mu\nu\rho}=\partial_{\mu}B_{\nu\rho} + \partial_{\rho}B_{\mu\nu}+ \partial_{\nu}B_{\rho\mu}$, $\tilde g$ and $c$ are arbitrary constants and we have included a dilaton-like coupling to the fields.
The Lagrangian (\ref{mass1}) describes a massless two-form field $B_{\mu\nu}$, with one degree of freedom, coupling to a massless gauge field $A_{\mu}$, with 2 degrees of freedom, plus the dilaton field. The fact that the two-form field has one degree of freedom is evident by the fact that in four dimensions, it transforms under the little group $SO(2)$. We will see that in its dual form, this Lagrangian describes a gauge field $A_{\mu}$ with 3 degrees of freedom, {\it i.e.}~the single degree of freedom carried by the two-form field is ``eaten'' by the gauge field to provide a mass, and the two-form field no longer appears.

To arrive at the dual form, we make an intermediate step which involves rewriting Eq.~(\ref{mass1}) by integrating the coupling term by parts, and then imposing the constraint $H=dB$ by way of a Lagrange multiplier field $\eta$. Integration by parts changes the form of the coupling term from $\epsilon^{\mu\nu\rho\sigma}B_{\mu\nu}F_{\rho\sigma}$ to $\epsilon^{\mu\nu\rho\sigma}H_{\mu\nu\rho} A_\sigma$, and so eliminates $B_{\mu\nu}$ from the Lagrangian. To retain the same information as was present in the original form, we need to impose the constraint $H=dB$, however, as we have already eliminated $B_{\mu\nu}$, we formulate the constraint in terms of the new field $H_{\mu\nu\rho}$ as $dH=0$ (which is of course true in the case that $H=dB$).
The Lagrangian in Eq.~(\ref{mass1}) can be thus rewritten as:
\be\label{mass2}
\mathcal{L}=-\frac{e^{\phi}}{12}H^{\mu\nu\rho}H_{\mu\nu\rho}
-\frac{e^{-\phi}}{4\,\tilde g^2}F^{\mu\nu}F_{\mu\nu}-\frac{c}{6}\epsilon^{\mu\nu\rho\sigma}H_{\mu\nu\rho}A_{\sigma}-\frac{c}{6}\, \eta \, \epsilon^{\mu\nu\rho\sigma} \partial_{\mu}H_{\nu\rho\sigma}\;.
\ee
Now, integrating by parts the last term in Eq.~(\ref{mass2}),
\be\label{mass3}
{\cal L} = -\frac{e^{\phi}}{12} H_{\mu\nu\lambda} H^{\mu\nu\lambda} -
\frac{e^{-\phi}}{4\, \tilde g^2} F_{\mu\nu} F^{\mu\nu} - \frac{c}{6} \epsilon^{\mu\nu\lambda\beta}H_{\mu\nu\lambda} (A_{\beta} + \partial_\beta \eta)
\ee
 and solving for $H$, we find:
$$ H^{\mu\nu\lambda} = -c \, e^{-\phi} \epsilon^{\mu\nu\lambda\beta}
(A_{\beta} + \partial_\beta \eta)\,.  $$
Inserting this back into Eq.~(\ref{mass3}), we find
\be
\mathcal{L}=-\frac{e^{-\phi}}{4\, \tilde g^2}F^{\mu\nu}F_{\mu\nu}
-\frac{c^2 e^{-\phi}}{2}(A_{\sigma}+\partial_{\sigma}\eta)^2.
\ee
Normalising  the kinetic term, we see that the gauge field $A_{\mu}$ has acquired a mass $m^2= \tilde  g^2c^2e^{-\phi}$.
Notice the the scalar $\eta$ can be gauged away  via a gauge transformation of $A \to A + \partial \Lambda$, thus we are left with only the mass term. By absorbing the scalar field $\eta$ into the gauge field, we have explicitly chosen a gauge.


In our set-up discussed in the main text, this mechanism is realised via the coupling of the gauge field $F_2$ to the RR two form $C_2$ (see Eq.~(\ref{wz3})). The kinetic term for $C_2$ descends from the 4D components of the RR field strength, $F_3 = dC_2$. This arises from the ten dimensional type IIB action. The relevant piece in the Einstein frame, is given by
\be
\frac{1}{2\kappa_{10}^2}\int d^{10}x \sqrt{-g_{10}}\left(-\frac{e^{\phi}}{12}
\,F_{\mu\nu\lambda} F^{\mu\nu\lambda}\right)\,,
\ee
where $2\kappa_{10}^2 = (2\pi)^7(\alpha')^4$ \cite{GKP}.
After dimensional reduction to four dimensions this becomes:
\be
\frac{M_P^2}{2}\int d^{4}x \sqrt{-g_{10}}\left(-\frac{e^{\phi}}{12}
\,F_{\mu\nu\lambda} F^{\mu\nu\lambda} \right)\,,
\ee
where we have used that
\be
\frac{1}{2 \kappa_{10}^2} \int d^6 x \,h \,\sqrt{-g_6}=\frac{V_6}{2\kappa_{10}^2}=\frac{M_P^2}{2}\,.
\ee
where $h$ is the warped factor.
Using this and (\ref{wz3}), we find the appropriate Lagrangian for the mass generation in our set-up,
\be
\mathcal{L}_{\rm mass}=-\frac{e^{\phi}}{12}\left(\frac{M_P^2}{2}\right)
F^{\mu\nu\rho}F_{\mu\nu\rho}-\frac{T_3 (2\pi\alpha')^2e^{-\phi}\,}{4}\,F^{\mu\nu}F_{\mu\nu}+\frac{T_3(2 \pi \alpha')}{4}\, \epsilon^{\mu\nu\rho\sigma}\mathcal{C}_{\mu\nu}F_{\rho\sigma}\;.
\ee
Rescaling the 2-form as, %
$
C_{2}=\frac{\sqrt{2}}{M_P}\widetilde{C}_2\;,
$
the Lagrangian takes the form of Eq.~(\ref{mass1})\footnote{Note that in our case, $C_2 = {\mathcal C}_2$.}
\bea
\mathcal{L}_{\rm mass} &=& -\frac{e^{\phi}}{12}\widetilde{F}^{\mu\nu\rho}\widetilde{F}_{\mu\nu\rho}-\frac{e^{-\phi}}{4\,\tilde g^2}F^{\mu\nu}F_{\mu\nu}+ \frac{c}{4}\, \epsilon^{\mu\nu\rho\sigma}\widetilde{C}_{\mu\nu}F_{\rho\sigma}\;,
\eea
where
\be
 \tilde g^{2}=\frac{1}{T_3(2\pi\alpha')^{2}}\,, \hspace{1cm}c=\frac{T_3 (2 \pi \alpha')\sqrt{2}}{M_P}\,.
 \ee

\ni Thus the dilaton dependent mass for the vector field is given by
\be
m^2 = \tilde g^2 c^2 e^{-\phi}= \frac{2\,T_3e^{-\phi}}{M_P^2}=\frac{(2\pi)^4M_s^2e^{-\phi}}{\mathcal{V}_6} \,,
\ee
where $M_s=\alpha'^{-1/2}$ is the string scale, $\mathcal{V}_6=V_6 / \ell_s^6$ is the dimensionless six dimensional volume and we have used that
\be
T_3 = (2 \pi)^{-3}(\alpha')^{-2}, \hspace{1cm}
M_P^2 = 2 (2 \pi)^{-7}\mathcal{V}_6 M_s^{2}\,.
\ee

Going back to the WZ action for the D3-brane in Eq.~(\ref{wz3}), using ${C}_4 = \sqrt{-g}\,  h^{-1}\, dx^0\wedge dx^1\wedge dx^2 \wedge dx^3$  and the mass term discussed above, we have

\be
S_{\rm WZ}= q \int  d^4x  \sqrt{-g}  \,\left(h^{-1}
-\frac{m^2}{2} \, {\mathcal A}_\mu  {\mathcal A}^\mu
+ \frac{{\mathcal C}_0 }{8} \, \epsilon^{\mu\nu\lambda\beta} {\mathcal F}_{\mu\nu}{\mathcal F}_{\lambda\beta}  \right),
\ee

\ni where here $ \tilde g {\mathcal A}_\mu = A_\mu $ is the canonical normalised gauge field (and correspondingly ${\mathcal F}_{\mu\nu} $ its field strength). Further,  $\epsilon^{\mu\nu\alpha\beta}$ is the Levi-Civita tensor, such that $\epsilon_{0123} = \sqrt{-g}$.

 %

\end{document}